\documentclass[preprint,aps,nofootinbib,floatfix,superscriptaddress,preprintnumbers]{revtex4}

\usepackage{psfrag}
\usepackage{amsmath,amssymb,graphicx,epsfig}
\usepackage{xspace,bm,slashed,color,verbatim}
\usepackage{feynmp}
\usepackage{subfigure}
\usepackage{color}
\usepackage[hyperfootnotes=false]{hyperref}
\usepackage{dsdshorthand}
\usepackage{bbm}

\usepackage{ifpdf}
\ifpdf
\else
\fi

\def\be{\begin{equation}}
\def\ee{\end{equation}}

\newcommand\cN{\mathcal{N}}

\newcommand\AdS{\mathrm{AdS}}
\newcommand\dS{\mathrm{dS}}
\newcommand\Li{\mathrm{Li}}
\newcommand\cW{\mathcal{W}}
\newcommand\cI{\mathcal{I}}
\renewcommand\e{\varepsilon}
\newcommand\rd{d}

\newcommand\fact{{\bf feature}\xspace}
\newcommand\facts{{\bf features}\xspace}

\newcommand\conf{\mathrm{}}

\newcommand{\OMIT}[1]{}

\newcommand{\eq}[1]{Eq.~(\ref{eq:#1})}

\renewcommand{\sec}[1]{Sec.~\ref{sec:#1}}

\newcommand{\app}[1]{App.~\ref{app:#1}}
\newcommand{\fig}[1]{Fig.~\ref{fig:#1}}

\newcommand{\nn}{\nonumber}

  \newcount\hour \newcount\minute
  \hour=\time \divide \hour by 60 \minute=\time
  \newcommand{\todaytime}{\today \ -- \number\hour :\ifnum \minute<10 0\fi\number\minute}

\begin{document}


\vbox{\hfill\hbox{MIT--CTP 4285}}

\title{\phantom{x}\vspace{1cm}\boldmath
Jet Physics from Static Charges in AdS
 \vspace{0.6cm}
}
\author{Yang-Ting Chien}
\affiliation{Center for the Fundamental Laws of Nature, Harvard University,\\
  Cambridge, MA~02138}

\author{Matthew D. Schwartz}
\affiliation{Center for the Fundamental Laws of Nature, Harvard University,\\
  Cambridge, MA~02138}

\author{David Simmons-Duffin}
\affiliation{Center for the Fundamental Laws of Nature, Harvard University,\\
  Cambridge, MA~02138}

\author{Iain W.~Stewart\vspace{0.5cm}}
\affiliation{Center for the Fundamental Laws of Nature, Harvard University,\\
  Cambridge, MA~02138}
\affiliation{Center for Theoretical Physics, Massachusetts Institute of
  Technology,\\ Cambridge, MA 02139\vspace{0.3cm}}


\begin{abstract}
  Soft interactions with high-energy jets are explored in radial coordinates
  which exploit the approximately conformal behavior of perturbative gauge
  theories. In these coordinates, the jets, approximated by Wilson lines, become
  static charges in Euclidean AdS. The anomalous dimension of the corresponding
  Wilson line operator is then determined by the potential energy of the
  charges.  To study these Wilson lines we introduce a ``conformal gauge'' which
  does not have kinetic mixing between radial and angular directions, and show
  that a number of properties of Wilson lines are reproduced through relatively
  simple calculations. For example, certain non-planar graphs involving multiple
  Wilson lines automatically vanish.  We also discuss the linear growth of the
  charges' imaginary potential energy with separation, and a relationship
  between Wilson line diagrams and Witten diagrams.
\end{abstract}
\maketitle
\thispagestyle{empty}


\setcounter{page}{1}

\section{Introduction}

The richness of quantum chromodynamics is hidden in its deceptively simple
Lagrangian $\mathcal{L} = - \frac{1}{4} F_{\mu \nu}^2 + \bar{q} i
D\!\!\!\!\slash\, q - m \bar{q} q$. At low energy, the theory has a mass gap
$\sim \Lambda_{\rm QCD}$ and a discrete set of bound states. At high temperature
it forms a quark-gluon plasma. At high energy, another phenomenon emerges: jets.
The preference for producing collimated jets arises from logarithmic enhancement
due to collinear and soft singularities. The cross section for production of
quarks alone is not infrared safe, but the cross-section for production of jets,
built from quarks accompanied by collinear and soft radiation, is a calculable
and well-defined quantity.

Another way to think about jets is through Sudakov logs. For example, consider
the mass of a jet $m_J$ computed in perturbation theory, assuming massless
quarks.  At leading order, the distribution is singular, $d\sigma/dm_J^2 \propto
\delta (m_J^2)$, since there is no radiation. At higher orders, the distribution
contains terms like $\alpha_s \frac{1}{m_J^2} \ln \frac{m_J^2}{Q^2}$ where $Q$
is a typical hard scale, like the jet energy. In terms of the integrated jet
mass $R (m_J^2) = \int_0^{m_J^2} dm'^2 (d \sigma/dm'^2)$, the series has the structure
\begin{equation}
  R (m_J^2) = 1 + \alpha_s \ln^2 \frac{m_J^2}{Q^2} + \alpha_s^2 \ln^4
  \frac{m_J^2}{Q^2} + \cdots
\end{equation}
The coefficients of these terms and the precise definition of $Q$ depend on the
particular process, and for simplicity, we have only shown the leading large
logarithms. These logs are Sudakov double logs, of the form $\alpha_s^n \ln^{2
  n} x$. They come from the region of overlapping soft and collinear divergences
and are present in any gauge theory with massless charged particles. Sudakov
logs invalidate the perturbation expansion. However when one re-sums the series,
the final non-perturbative expression, schematically $R (m_J^2) = \exp (-
\alpha_s \ln^2 \frac{m_J^2}{Q^2})$ vanishes at $m_J^2 = 0$ implying that the
cross section for producing massless quarks is zero. The objects that are
produced are jets, of finite mass. The coefficient of the Sudakov log in this
exponential is a function of the coupling constant $\Gamma_{\mathrm{cusp}}
(\alpha_s)$ called the cusp anomalous dimension.

Sudakov logs and the cusp anomalous dimension are simplest to study in the soft
limit of QCD, where one treats a massless parton (quark or gluon) as a hard
charged object plowing through a background of soft radiation. The soft
radiation cannot change the direction or energy of the hard parton, and so the
parton factorizes out as a Wilson line source for soft gluons. This treatment of
soft radiation becomes manifest when using QCD factorization theorems to
describe hard collisions, see the reviews~\cite{Collins:1989gx,Sterman:1995fz},
or when using soft-collinear effective theory
(SCET)~\cite{Bauer:2000ew,Bauer:2000yr,Bauer:2001ct,Bauer:2001yt,Bauer:2002nz}
to describe the interaction of soft and collinear partons in hard collisions.
Thus using Wilson line operators the soft interaction properties of jets can be
investigated. Wilson lines also appear in the study of scattering in planar
$\cN=4$ SYM, via a surprising duality relating null polygonal loops to
scattering amplitudes
\cite{Alday:2007hr,Drummond:2007aua,Brandhuber:2007yx,Drummond:2007cf,Drummond:2007bm,Bern:2008ap,Drummond:2008aq,Berkovits:2008ic}.

A Wilson line is defined as
\begin{equation} \label{njet}
  \mathcal{W} (C) = \mathcal{P}  \exp \p{ i g \int_C A_{\mu} \rd
  x^{\mu} } \,,
\end{equation}
where $C$ is a contour describing the path of the partons and $\mathcal{P}$
denotes path-ordering, along the contour. Typically, one takes $C$ to be a
simple closed contour, and makes $\cW$ gauge invariant by taking a trace.  We
will be creating gauge invariants using $\cW$ in more complicated ways.  For
example, a process like $e^+ e^- \rightarrow \mathrm{hadrons}$ is, to leading
order in $\alpha_s$, described by $e^+ e^- \rightarrow \bar{q} q$ with the
quarks traveling off in the $n_1^{\mu} = (1, \vec{v})$ and $n_2^{\mu} = (1, -
\vec{v})$ directions.  In this case, we can write $\cW(C)$ as the product of two
Wilson lines from $0$ to $\infty$ along $n_1$ and $n_2$, with one in the
fundamental and one in the anti-fundamental representation.

More generally, for $N$-jet production
in $e^+e^-$ collisions or $(N-2)$-jet production in $pp$ or $p\bar p$
collisions, we are interested in a product of $N$ Wilson lines along directions
$n_i^{\mu}$,
\begin{align}\label{Wilsonlinewithtensor}
  \cW_{d_1, \cdots, d_N}(n_1,\cdots,n_N)  &= t_{c_1,\dots,c_N}  \prod_{i=1}^N
    \p{\, 
  {\mbox{\large $\cP$}} 
  \exp\, ig \int_0^\oo\! ds\ n_i\cdot\! A^a(s\, n_i)\bm{T}^a_i}^{c_i}_{d_i}
  \,.
\end{align}
The lines here are all outgoing. For an incoming Wilson line we simply replace
the path-ordering, $\cP$, by anti-path ordering, $\overline\cP$, and replace the
$ig$ with $-ig$.  Let us take a moment to explain the remaining notation.  The
$\bm{T}_i$ are gauge generators in the color representation $R_i$ associated
with parton $i$. They satisfy $[\bm{T}_i,\bm{T}_j]=0$ for $i\neq j$, along with
the color conservation relation $\sum_i \bm{T}_i=0$. For light quarks and gluons
the directions $n_i^{\mu}$ are light-like, $n_i^2=0$, while for heavy quarks
like the top where mass effects are important, we have $n_i^2 \neq 0$. To keep
our discussion general we will mostly work with $n_i^2\ne 0$. Often in the
literature the time-like component of $n_i^\mu$ is taken to be positive, and the
integration along the path extends from $s=0$ to $\infty$ for outgoing partons
and from $s=-\infty$ to $0$ for incoming partons.  For simplicity we will always
take $s=0$ to $\infty$ and let $n_i^\mu$ have a negative time-like component for
incoming particles.\footnote{Another common convention in the literature is to
  use tangent vectors to the contour $v_i^\mu$, in place of our $n_i^\mu$. For a
  $2$-jet Wilson line for $e^+e^-\to q\bar q$, the relation to our conventions
  is $v_1=-n_1$ to $v_2=n_2$, where $n_1$ and $n_2$ have positive
  time-components.}

$\cW$ depends on a tensor $t_{c_1,\ldots,c_N}$, where the $c_i$
denote the color indices at $s=0$.  These tensors live in the color-invariant
subspace $\cI$ of the tensor product of representations associated with each
jet
\begin{align}  \label{I}
  \cI = \Big( R_1\otimes R_2\otimes\dots\otimes R_N \Big)_{\text{color singlet
      subspace}} \,.
\end{align}
In a scattering process, the short-distance physics at the origin specifies the
relevant channels and determines the $t_{c_1,\ldots,c_N}$, which are
Clebsch-Gordan coefficients.

The $d_i$ indices on Eq.~(\ref{Wilsonlinewithtensor}) denote the color indices
at $s=\infty$.  Matrix elements of $\cW(n_i)$ will be infrared divergent unless
the $d_i$ are contracted, as in various physical calculations. For example, 
Wilson line matrix elements contribute to matching calculations in SCET, see
eg.~\cite{Bauer:2000yr,Manohar:2003vb,Bauer:2006mk,Bauer:2006qp,Fleming:2007xt,Kelley:2010fn}.
Operators describing the hard interaction for $N$-jet production appear in the
SCET Lagrangian as
\begin{align} \label{eq:SCETO}
  {\cal L} &= C_{c_1,\ldots,c_N}(s_{ij}) {\cal O}^{c_1,\ldots,c_N}(n_i) \nn\\
 &= C(s_{ij}) \big[ \chi_{n_j}^{d_j} \cdots \bar \chi_{n_k}^{d_k}
  \cdots {\cal B}^{\perp d_\ell}_{n_\ell}\ldots \big]\, \cW_{d_j,\cdots,d_k,\cdots,d_\ell,\cdots}(n_i) \,,
\end{align}
where $C_{c_1,\ldots,c_N}(s_{ij}) = t_{c_1,\ldots,c_N}C(s_{ij})$ is a
Wilson coefficient depending on hard scales $s_{ij}=p_i \cdot p_j$, where $p_i =
n_i Q_i$ are the jet four-momenta at leading power (and in general we have a sum
over terms of this sort for the possible color structures $t_{c_1,\ldots,c_N}$).
In square brackets are collinear quark ($\chi_{n_j}$) and gluon (${\cal
  B}^{\perp}_{n_\ell}$) fields that are each contracted in color with the $d_j$
or $d_\ell$ indices from the Wilson line. In a matching computation the infrared
divergences from matrix elements of collinear fields and from the soft Wilson
lines combine to yield the same infrared divergences as for the corresponding
matrix element in QCD, ensuring that the Wilson coefficients $C(s_{ij})$ are
finite. In this computation there is a cancellation of overlapping infrared and
ultraviolet divergences between the collinear matrix elements and matrix
elements involving the soft Wilson lines.

Alternatively, the Wilson line can be used to calculate a soft function (for
examples
see~\cite{Korchemsky:1997sy,Fleming:2007xt,Schwartz:2007ib,Bauer:2008dt,Becher:2009qa,Becher:2009th,Kelley:2010qs,Ellis:2010rwa,Jouttenus:2011wh}),
which appear in physical cross sections for hard processes with jet production,
\begin{align} \label{eq:S}
 S(k) = \langle 0 | \cW_{d_1,\cdots, d_N}(n_i) \hat
  M(k) \cW^\dagger_{d_1,\cdots, d_N}(n_i) | 0 \rangle \,.
\end{align}
In this case, the $d_i$ indices of the Wilson line are contracted with those of
its adjoint and the product includes a measurement function $M(k)$ which acts on
final state soft partons, measuring momentum components $k$. These soft
functions are cross sections for soft radiation and are infrared finite by
themselves.

An important property of Wilson lines is that, even though they are non-local
objects, they are multiplicatively renormalizable.  A number of \facts are known
(or conjectured) about the renormalization of $\cW(C)$.
\begin{enumerate}
\item If the contour $C$ is smooth and not self-intersecting, any ultraviolet
  divergences in correlators of $\cW(C)$ are exactly canceled by field strength
  and coupling constant counterterms.\footnote{At least this is the case in
    dimensional regularization.  More generally, there can be an overall linear
    divergence proportional to the length of the Wilson line, which can also be
    subtracted off with an appropriate counterterm.} For the $\mathcal{W} (n_i)$
  relevant for jet physics, the Wilson line has kinks and self-intersections in
  its path at the origin, and $C$ is not smooth. In this case additional
  divergences are present and the Wilson line picks up an anomalous dimension
  $\G$. This anomalous dimension can only depend on the angles $\beta_{ij}$
  where the contour abruptly changes direction or on crossing angles at
  self-intersections~\cite{Brandt:1981kf}.  When the direction of the contour
  changes from $n_i^\mu$ to $n_j^\mu$, the {\it cusp angle} is
 \begin{equation}
    \cosh \beta_{ij} = \frac{n_i\.n_j}{|n_i||n_j|} \,,
    \label{eq:betadefinition}
  \end{equation}
  where here we consider paths where $|n_i|^2=n_i^2\ne 0$.  For two jets from
  $e^+e^-\to q\bar q$ (with massive quarks), $\b_{12}$ is real.  More generally, for $e^+ e^-
  \rightarrow N\ \mathrm{jets}$, all the $n_i^{\mu}$ correspond to final state
  jets and each $\beta_{ij}$ is real.

  \item At order $\alpha_s$, the anomalous dimension is
  \begin{equation}
    \Gamma = -\frac{\alpha_s}{\pi} \sum_{i<j} \bm{T}_i\.\bm{T}_j
   \left( (\beta_{i j} - i \pi) \coth \beta_{i j} - 1 \right)
   \label{eq:1loopcuspintermsofbeta}
  \end{equation}
  where the sum is over pairs of jet directions $n_i$, $n_j$.  The color
  structure $\bm{T}_i\.\bm{T}_j\equiv \bm{T}^a_i\bm{T}^a_j$ involves the
  generators from Eq.~(\ref{Wilsonlinewithtensor}).  These generators allow
  the anomalous dimension $\G$ to mix the different invariant tensors $t_{c_i}$ in
  Eq.~(\ref{Wilsonlinewithtensor}) during renormalization group flow.
  $\G$ is an operator on the space $\cI$, which we can write as a general
  expression with generators $\bm{T}_i$ acting on the $i$-th tensor factor.

\item In situations like jet production in hadron collisions or deep inelastic
  scattering (DIS), there are both initial state and final state Wilson lines.
  Initial state Wilson lines follow paths that extend backward in time from the
  origin, $n^{\mu} = (- 1, \vec{v})$.  For cusps between initial and final state
  partons $\beta_{ij}$ is complex, but we can define a real cusp angle
  $\gamma_{ij}= \beta_{ij} - i \pi$. This alternative definition of the cusp
angle is related to the previous one by a sign
  \begin{equation}
    \label{eq:betagamma}
    \cosh \gamma_{i j} = - \frac{n_i \cdot n_j}{|n_i||n_j|}
                       = - \cosh \beta_{i j} \,,
  \end{equation}
  The angles $\beta_{ij}$ and $\gamma_{ij}$ are illustrated in
  Fig.~\ref{fig:betagamma}.  Whether $\gamma_{ij}$ or $\beta_{ij}$ are complex
  affects the complexity of the anomalous dimension, which has physical
  consequences.  For example, these factors of $i \pi$ can partly explain the
  large $K$-factor in the Higgs production cross section~\cite{Ahrens:2008qu}.

  \begin{figure}[t!]
\subfigure[\ $e^+e^-$ to dijets]{\label{fig:dijetangles}
\begin{fmffile}{dijetangles}
\begin{fmfgraph*}(90,90)
\fmfset{arrow_len}{3mm}
\fmfsurroundn{v}{12}
\fmfv{label=$n_2$}{v3}
\fmfv{label=$n_1$}{v5}
\fmf{plain}{center,u3}
\fmf{fermion}{u3,v3}
\fmf{plain}{center,u5}
\fmf{fermion}{u5,v5}
\fmf{phantom}{center,u9}
\fmf{phantom}{u9,v9}
\fmf{phantom}{center,u11}
\fmf{phantom}{u11,v11}
\fmffreeze
\fmf{dots,right=.3,label=$\beta_{12}$}{u3,u5}
\end{fmfgraph*}
\end{fmffile}
}
\qquad
\subfigure[\ DIS]{\label{fig:DISangles}
\begin{fmffile}{DISangles}
\begin{fmfgraph*}(90,90)
\fmfset{arrow_len}{3mm}
\fmfsurroundn{v}{12}
\fmfv{label=$n_2$}{v3}
\fmfv{label=$n_1$}{v11}
\fmfv{label=$-n_1$}{v5}
\fmf{plain}{center,u3}
\fmf{fermion}{u3,v3}
\fmf{dashes}{center,u5}
\fmf{dashes_arrow}{u5,v5}
\fmf{phantom}{center,u9}
\fmf{phantom}{u9,v9}
\fmf{plain}{center,u11}
\fmf{fermion}{u11,v11}
\fmffreeze
\fmf{dots,right=.3,label=$\gamma_{12}$}{u3,u5}
\end{fmfgraph*}
\end{fmffile}
}
\caption{ Our definitions for the cusp angles $\beta_{12}$ and $\gamma_{12}$.
  $\beta_{12}$ is real when both Wilson lines represent final-state partons,
  while $\gamma_{12}$ is real when one Wilson line represents a final state
  parton, and one represents an initial state parton.}
\label{fig:betagamma}
\end{figure}
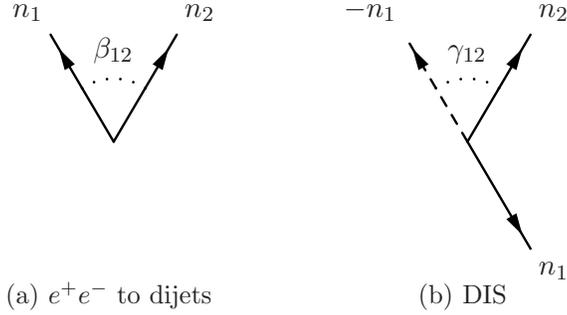

  \item {\it Abelian exponentiation}: The anomalous dimension in QED
  without propagating quarks is one-loop exact.

\item In the limit that the tangent vectors become lightlike, $n_i^2 \rightarrow
  0$ and $\beta_{ij} \rightarrow \infty$, the anomalous dimension becomes
  {\it linear} in the cusp angles $\beta_{i j}$ to all orders in perturbation
  theory~\cite{Brandt:1981kf, Korchemsky:1987wg, Alday:2007mf},
   \begin{equation} \label{eq:linearbeta}
     \Gamma = -\sum_{i<j} {\mathbf{\Gamma}}^{i j} (\alpha_s)
 \beta_{i j}+\dots
   \end{equation}
   where ``$\dots$" are terms that are constant or go to zero as
   $\b_{ij}\to\oo$.  The possibility that $\mathbf{\Gamma}^{ij} (\alpha_s)=
   \bm{T}_i \cdot \bm{T}_j \Gamma_{\rm cusp}(\alpha_s)$, where $\Gamma_{\rm
     cusp}(\alpha_s)$ is independent of $i$ and $j$, is known as {\it Casimir
     scaling}. This has been shown by explicit calculations for two lines at
   three loops~\cite{Vogt:2004mw}, and by other arguments for multiple lines up
   to 4-loops~\cite{Becher:2009qa}.

   ~~~~Exactly at $n_i^2 = 0$, $\beta_{ij} = \infty$ and the anomalous
   dimension is singular. This singularity is due to additional overlapping soft
   and collinear divergences, and induces dependence of the anomalous dimension
   on the renormalization group scale $\mu$.  For example, at one loop,
  \begin{equation} \label{eq:onelooplight}
    \Gamma = \frac{\alpha_s}{\pi} \sum_{i<j} \bm{T}_i \cdot
    \bm{T}_j \ln \frac{\mu^2}{n_i\cdot n_j \Lambda^2} + \dots \,.
  \end{equation}
  Here $\Lambda$ is another scale with dimension of mass. In matrix elements of
  Wilson lines, which are infrared divergent, $\Lambda$ is related to the
  infrared regulator.  When collinear graphs are included in the the calculation
  of Wilson coefficients for a hard scattering process, as in
  Eq.~(\ref{eq:SCETO}), the infrared regulator will cancel and $\Lambda$ will be
  replaced by a hard scale $\Lambda^2 \to \pm Q_i Q_j$.  In calculating soft
  functions, as in Eq. \eqref{eq:S}, the infrared divergences will cancel
  between real and virtual contributions, and the scale will be replaced by a
  physical one, relevant to the soft function $\Lambda^2 \to k_ik_j$.  At all
  orders, the anomalous dimension is linear in $\ln \mu^2$, for the same reason
  that $\Gamma$ is linear in $\beta_{i j}$ at large cusp angles.  Proofs of
  \eq{linearbeta} for two Wilson lines have been given in
  Refs.~\cite{Korchemsky:1987wg,Manohar:2003vb,Bauer:2003pi}, and for multiple
  lines in Refs.~\cite{Chiu:2008vv,Becher:2009qa,Chiu:2009mg}.

\item At 1-loop the anomalous dimension must be a sum over pairs of Wilson
  lines. Surprisingly some pairwise structure seems to persist to higher orders
  in perturbation theory. For example, in the massless case, the anomalous
  dimension of a 4-jet Wilson line at 2-loops, as a matrix in color space, was
  found to be exactly proportional to the 1-loop anomalous dimension, an
  unexpected result~\cite{Aybat:2006wq}.

 ~~~~ It has been conjectured that for massless jets the anomalous dimension of
  the $N$-jet Wilson line has a pairwise structure to all orders in perturbation
  theory for the $\ln \mu$ term~\cite{Becher:2009cu,Chiu:2009mg,Gardi:2009qi,Becher:2009qa,Bern:2008pv}
  \begin{equation}
  \label{eq:neubertgardiconjecture}
    \Gamma \stackrel{?}{=}  \sum_{i \neq j} \Gamma^{i j}_{\mathrm{cusp}} (\alpha_s)
    \bm{T}_i \cdot \bm{T}_j \ln \frac{\mu^2}{n_i\cdot n_j \Lambda^2} +
    \gamma(\alpha_s,\{n_k\cdot n_\ell \}) \,.
  \end{equation}
  Given \eq{linearbeta}, \eq{neubertgardiconjecture} becomes non-trivial for
  four or more Wilson lines where matrices appear for the color structures.  If
  there were a general proof of Casimir scaling it would imply that
  $\Gamma^{ij}_{\mathrm{cusp}} (\alpha_s)$ cannot depend on the representations
  $i$ and $j$, making the coefficient a universal function
  $\Gamma^{ij}_{\mathrm{cusp}} (\alpha_s)=\Gamma_{\mathrm{cusp}}(\alpha_s)$.
  This was conjectured in~\cite{Becher:2009qa,Becher:2009cu}.

  It has been furthermore conjectured that the regular anomalous dimension
  $\gamma$ is independent of conformal cross ratios (combinations of cusp angles
  $\beta_{ij}+\beta_{k\ell}-\beta_{ik}-\beta_{j\ell}$ that approach nontrivial
  constants as the $\b_{ij}\to\oo$), so that~\cite{Becher:2009qa}
 \begin{align}
  \gamma(\alpha_s,\{n_i\cdot n_j\}) \stackrel{?}{=} \sum_i \gamma^i (\alpha_s) \,.
 \end{align}
 This is known to be true to ${\cal O}(\alpha_s^2)$. At ${\cal O}(\alpha_s^3)$
 and beyond general constraints on the form of $\gamma(\alpha_s,\{n_i\cdot
 n_j\})$ were reviewed in Ref.~\cite{Dixon:2010zz,Becher:2009qa}. Dependence on
 conformal cross ratios appears not to be forbidden by symmetry arguments, but
 whether this dependence exists is an open question. Possible terms at ${\cal
   O}(\alpha_s^3)$ which were not obviously forbidden were suggested in
 Ref.~\cite{Dixon:2010zz}. Very recently it was argued that these terms are
 forbidden by considerations from the Regge limit in
 Refs.~\cite{DelDuca:2011ae,DelDuca:2011xm}.
\end{enumerate}
Most of these results have been shown only through direct, and sometimes laborious
calculations. Even a simple result, such as Abelian exponentiation, requires
the use of eikonal identities and monitoring of combinatoric
factors. In this paper, we will show how some of these results can be understood in a
simple way using a mapping inspired by the approximate conformal invariance of
QCD.

At the classical level, QCD is conformally invariant. This symmetry is broken by
quantum effects, but for high energy scattering it continues to have
implications for the structure of perturbative results. Examples of the
implications of conformal symmetry for QCD can be found in
Refs.~\cite{Belitsky:2003ys,Braun:2003rp,Lipatov:1993yb,Faddeev:1994zg,Korchemsky:1994um,Alday:2007mf}.
Our main focus here will be on exploiting conformal invariance to understand
properties of the anomalous dimensions of Wilson lines relevant for jet physics.

\begin{figure}[t!]
\includegraphics[width=\textwidth]{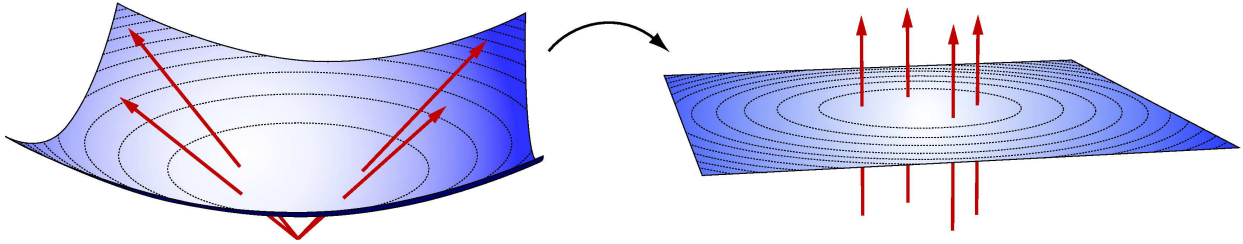}
\caption{ A coordinate change maps Minkowski space to $\R\x\AdS$.  In this
  figure the outgoing Wilson lines become static charges in AdS, and their tree
  level energy in AdS is equal to the original one-loop anomalous dimension for
  the lines.  }
\label{fig:radialquantization}
\end{figure}

A Wilson line emanating from the origin in the direction $n^{\mu}$ comprises the
points $x^{\mu} = s n^{\mu}$, for $s>0$. A scale transformation is simply a
change in $s$.  Scale invariance is made more manifest by defining a new time
coordinate $\tau \equiv \ln |x|$ where $|x|$ is the (Lorentzian) distance from
the origin.  This makes the Wilson lines parallel, as shown in
Figure~\ref{fig:radialquantization}.  In terms of $\tau$, rescaling becomes
simply time translation, and conformal symmetry becomes the statement that the
physics is time translation invariant in $\tau$.  The Wilson lines become static
charges whose energy is the anomalous dimension. Spatial slices in these
coordinates are copies of Euclidean Anti-deSitter space (AdS).  In this paper,
we describe how many of the features of Wilson lines enumerated above can be
understood in AdS coordinates. For example, that the Coulomb potential is
one-loop exact in QED automatically implies that the anomalous dimension of
multiple Wilson line operators are one-loop exact in QED.

In addition to providing a pleasing physical picture, radial coordinates make a
number of calculations much easier. One of the reasons that the classical
conformal invariance of QCD rarely simplifies diagrammatic computations is that
it is broken by standard gauge choices, such as Feynman gauge. Feynman gauge in
flat space leads to kinetic mixing between the time-like components $A_{\tau}$
and space-like components $A_i$ of the gauge field in $\R\x\AdS$.  Here we introduce a
new gauge, which we call {\it conformal gauge}, in which there is no such
mixing. One consequence is that in this gauge, since only $A_{\tau}$ is sourced
by the Wilson lines, and there is no $A_{\tau}^3$ or $A_{\tau}^4$ vertex in
Yang-Mills theory, many of the non-planar graphs at 2 and 3-loops automatically
vanish.  This automatically implies that the only graphs at 2-loops contributing
to the anomalous dimension have gluons going between pairs of Wilson lines,
which strongly suggests a pairwise structure.  This reasoning alone does not imply
that $\G$ must be pairwise at 3-loops or higher.

The organization of this paper is as follows. In \sec{conformal} we discuss the
appropriate mapping of Minkowski to AdS coordinates and simple implications for
multi-Wilson line configurations. In \sec{ClassicalAdSEnergies} we compute the
one-loop anomalous dimension of $\cW$ by carrying out a classical energy
computation in AdS, paying special attention to boundary conditions and
differences for incoming and outgoing lines.  We repeat this computation using
the standard one-loop diagrams in \sec{oneloop}, but utilizing the AdS
coordinates. A discussion of the lightlike limit and a way to think about the imaginary
energy in AdS is given in \sec{lightlikelimit}. In \sec{conformalgauge}, we introduce conformal gauges
which do not mix the time and spatial components of the gauge boson propagator
in $\R\x\AdS$, and in \sec{2loops} we demonstrate the utility of such gauges by
computing a two-loop contribution to the anomalous dimension of $\cW$ in a
simple way.  In \sec{wittendiagrams}, we mention an interesting formal relation
between anomalous dimension calculations for $\cW$ and Witten diagrams.  We
conclude in \sec{conclusion}.  Several technical discussions are relegated to
appendices.  In \app{genconformal} we construct the most general class of conformal gauges
without auxiliary parameters, and in \app{ghost} we give the corresponding
Feynman rules for ghosts.

\section{Conformal Coordinates}  \label{sec:conformal}

Having replaced hard partons by Wilson lines, the soft physics is described
simply by Yang-Mills theory, which enjoys classical conformal invariance in four
dimensions.  Equivalently, the soft action coupled to a background metric
$g_{\mu\nu}$ is Weyl-invariant: it is unchanged under a local rescaling of the
metric, $S_\mathrm{YM}[g]=S_\mathrm{YM}[e^{2\w(x)}g]$.  Although this symmetry
is broken by fermion masses and the QCD scale anomaly that generates
$\Lambda_{\rm QCD}$, it continues to have important implications for scattering
amplitudes at high energies.

One reason conformal invariance can be useful for QCD is that
some quantities are completely insensitive to the breaking of conformal invariance.  For example, the one-loop
cusp anomalous dimension Eq.~(\ref{eq:1loopcuspintermsofbeta}) is independent of
the matter content of the theory, since Feynman diagrams contributing to it only
involve a single gluon exchanged between Wilson lines.  Consequently, it has a
universal form, and we can compute it assuming exact
conformal invariance. In other words, we can compute it in our favorite
  conformal theory, for instance $\cN=4$ SYM, and the result will hold in any
  gauge theory.  In the next two sections, we will use this fact to give a
simple and intuitive derivation of Eq.~(\ref{eq:1loopcuspintermsofbeta}).

To the extent that conformal symmetry is a good approximation, it is natural to
apply techniques which have proved useful for studying conformal field theories
in other contexts. In particular, we consider {\it radial quantization} around
the origin. In Ref.~\cite{Alday:2007hr} this was used to study the anomalous
dimensions of high spin operators, and has also been used in
Ref.~\cite{Belitsky:2003ys}. In this section, we review the relevant ideas in
the context of $N$-jet Wilson lines.

Consider a Wilson line in the direction $n^\mu$.
We can write $n^\mu = (\cosh\b,\sinh\b\,\hat\bn)$, with $\hat \bn$ a unit vector in $\R^3$, and $x^\mu=e^\tau n^\mu$.
The path of the Wilson line is then described by
\begin{align}
\label{eq:pathinradialcoordinates}
  t = e^{\tau} \cosh \beta,
  \qquad
  r = e^{\tau} \sinh \beta,
  \qquad
  \b,\th,\f\ \textrm{fixed},
\end{align}
with $\tau$ running from $-\oo$ to $\oo$.  Wilson lines in different directions
will correspond to different values of $\beta, \theta$ and $\phi$.  In these
coordinates, the Minkowski metric becomes
\begin{align}
  d s^2_{\R^{1,3}} &= d t^2 - d r^2 - r^2 d \Omega_2^2 \nn\\
  &=e^{2 \tau} \left[ d \tau^2 - (d \beta^2 + \sinh^2\beta \, d\Omega_2^2) \right].
  \label{eq:timedependentradialmetric}
\end{align}
Strictly speaking, this metric describes only a patch of $\R^{1,3}$ --- namely
the interior of the future light-cone.  We will return to this point shortly.

The idea of radial quantization is to interpret $\tau$ as a new time coordinate.
A na\"ive complication in this picture is that the metric
Eq.~(\ref{eq:timedependentradialmetric}) is now time-dependent.  However, in a
conformal theory, the dynamics is independent of the local scale, and we can
equivalently consider our theory with any metric related via $d s^2\rightarrow
e^{2 \omega (x)} d s^2$.  Thus, let us drop the overall $e^{2\tau}$ to obtain a
simple time translation-invariant product space,
\be
\label{eq:RcrossAdSmetric}
d s^2_{\R\x\AdS} = d \tau^2 - (d \beta^2 + \sinh^2\beta \, d\Omega_2^2).
\ee
The spatial part of this metric is the 3D hyperboloid, or Euclidean
Anti-deSitter space. With a slight abuse of nomenclature, we call it simply AdS.

In radial coordinates, the origin maps to $\tau = - \infty$ and motion along a
Wilson line corresponds to shifts in $\tau$. So to the extent that our theory
was scale invariant in Minkowski space, it is now time-translation invariant in
$\R\x\AdS$. Each Wilson line sits at fixed $(\b,\th,\f)$ and extends from $-\oo$
to $\oo$ in the time coordinate $\tau$.  That is, each Wilson line becomes a
static charge in AdS.  For perturbative computations in QCD (where conformal
invariance is broken by the scale anomaly) we may simply adopt the change of
coordinates in \eq{pathinradialcoordinates} as a method to carry out
computations. If the computation involves ingredients satisfying the conformal
invariance then the factors of $e^\tau$ will cancel out, and the result will be
constrained by properties of the $\AdS$ space.

For the sake of doing calculations, a key point is that the dilatation operator in
Minkowski space maps to the Hamiltonian in AdS in radial quantization,
\begin{equation}
  {\mathcal D}^{\mathrm{\R^{1,3}}} = x^{\mu} \partial_{\mu} = \partial_{\tau} =
  i\mathcal{H}^{\mathrm{\R\x\AdS}}.
  \label{eq:matchingofdilatationandhamiltonian}
\end{equation}
Consequently, the eigenvalue of dilatation --- the dimension (or when acting on
classically scale invariant Wilson lines, the anomalous dimension) --- is just
$i$ times the energy in AdS. So we can calculate anomalous dimensions by
calculating energies and apply our intuition from
electrodynamics to understand anomalous dimensions of Wilson lines.

What can the energy of two static charges in AdS depend on? Since the space is
homogeneous, it can depend only on the geodesic distance between the charges.
For example, suppose we have two time-like Wilson lines, pointing in the
directions $n_1^{\mu}$ and $n_2^{\mu}$, normalized so that $n_1^2 = n_2^2 = 1$. We
may first go to the rest frame of one, $n_1^{\mu} = (1, 0, 0, 0)$, and then
rotate so the other is at $n_2^{\mu} = (\cosh \beta_{12}, \sinh \beta_{12}, 0,
0)$. Then the geodesic distance between them, using the spatial part of the
metric Eq.~(\ref{eq:RcrossAdSmetric}), is just
\begin{equation}
  \Delta s = \beta_{12} \,.
\end{equation}
Considering also that $\frac{n_1 \cdot n_2}{|n_1||n_2|} = \cosh \beta_{12}$, we
see that the geodesic distance in AdS is the cusp angle. Thus, the energy of the
two charges, and hence the anomalous dimension in Minkowski space, can depend
only on the cusp angle. This was \fact~{\bf 1} from the introduction.
More
succinctly, the original Lorentz symmetry of Minkowski space becomes the
isometry group of AdS in radial coordinates.  Just as Lorentz invariance
dictates that the anomalous dimension can depend only on the cusp angle, the
isometries of AdS dictate that the energy can depend only on the geodesic
distance.

\begin{figure}[t!]
\begin{center}
\begin{psfrags}
\psfrag{b}[B][B][1][0]{$\AdS_3,\ \b\in\R$}
\psfrag{g}[B][B][1][0]{$\AdS_3,\ \g\in\R$}
\psfrag{d}[B][B][1][0]{$\mathrm{dS}_3$}
\psfrag{i}[B][B][1][0]{initial state}
\psfrag{f}[B][B][1][0]{final state}
\psfrag{z}[B][B][1][0]{$\ \ x=0$}
\includegraphics[width=60mm]{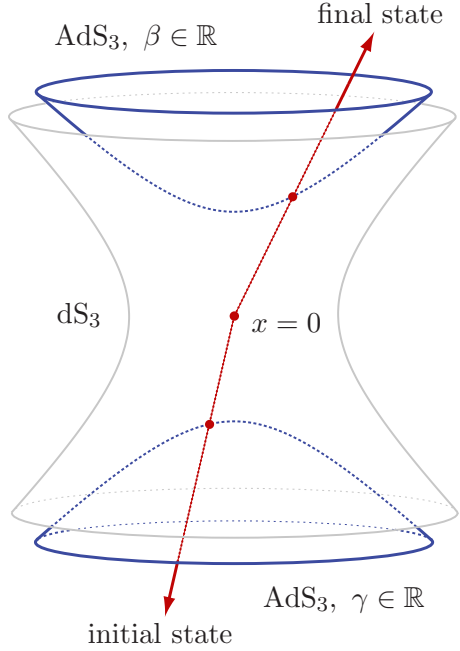}
\end{psfrags}
\end{center}
\caption{In radial quantization, final state lines map to a copy of $\AdS_3$ at positive Minkowski times, while initial state lines map to a second copy of $\AdS_3$ at negative Minkowski times.  Points that are spacelike separated from the origin map to $\dS_3$.}
\label{fig:copiesofAdS}
\end{figure}

We can also consider initial state Wilson lines. For example, in deep inelastic
scattering as Bjorken $x \to 1$, the initial state contains an energetic proton
in the Breit frame, the final state contains a jet, and the Wilson line
description applies. In our convention, the spatial vectors for these lines
still point out from the origin, and the lines extend to negative Minkowski
times.  Instead of Eq.~(\ref{eq:pathinradialcoordinates}), the path of an
initial state line is then described by
\begin{align}
t = -e^{\tau} \cosh\g,
\qquad
r = e^{\tau} \sinh\g,
\end{align}
for fixed real $\g$. When comparing the coordinates for parallel initial and
final state lines we have $\hat n\to -\hat n$ so $\theta\to\theta+\pi$ and
$\phi\to \phi+\pi$.  In radial quantization, initial state lines map to static
charges in a different copy of $\R\x\AdS$ comprising points in the interior of
the past light-cone (Figure~\ref{fig:copiesofAdS}). It is useful to think of
this second copy of AdS as being related by analytic continuation to the first.
Since from Eq.~(\ref{eq:betagamma}) we have $\cosh \g=-\cosh\b$, we can write
$\b=\g+i\pi$.  Both copies of AdS (along with a copy of deSitter space
describing points at spacelike separation from the origin) are related by
analytic continuation to the three-sphere $S^3$ that one would obtain by
repeating the exercise of radial quantization starting from Euclidean space,
$\R^4$.  This will be a useful tool in the following section.

For most of the remainder of the paper, we will focus on time-like Wilson lines
whose directions are normalized to $n^2 = 1$. Since all the energies and
dimensions are independent of rescaling of the $n$'s, the dependence on $|n|$
can be put back by dimensional analysis: $n^{\mu}_i \rightarrow
\frac{n_i^{\mu}}{|n_i |}$.  We will also have occasion to consider the light-like
limit $n^2\to 0$, which is phenomenologically relevant for the majority of
processes at colliders. Many of the properties of the light-like case can be
derived as a limiting case of the general time-like results. In the light-like
limit $n_i^2 \rightarrow 0$, and the charges move towards the boundary of AdS with
$\beta\to \infty$. Some results simplify for $n^2 = 0$, and when appropriate we
will consider this case separately.

\section{Classical AdS energies}
\label{sec:ClassicalAdSEnergies}

In radial coordinates, we have seen that the anomalous dimension of  a
collection of Wilson lines intersecting at a point is proportional to the energy of a
collection of static charges in AdS. This energy can only depend on the
geodesic distance between the charges, which is the same as the cusp angle
$\beta_{i j} = \cosh^{- 1} n_i \cdot n_j$. Now let us calculate that energy.

The energy of two charges in QCD at leading order is given, as in QED, by
solving Laplace's equation for the scalar potential $A_{\tau}$
in the presence of point sources $J_{\mu}$ given by $J_{\tau} = \delta^3 (x)$ and $\vec{J}=0$. The
homogeneous solutions are
\begin{equation}
  \frac{1}{\sinh^2 \beta} \partial_{\beta} \left( \sinh^2 \beta \left(
  \partial_{\beta} A_{\tau} \right) \right) = 0 \hspace{1em} \Rightarrow
  \hspace{1em} A_{\tau} (\beta) = C_1 + C_2 \coth \beta \label{wrong}
\end{equation}
Unfortunately, neither of these is the physically correct answer. This can be
seen most easily by looking at the large $\beta$ limit, where we expect
$A_{\tau} (\beta)$ to be linear in $\beta$. In this limit Eq.~(\ref{wrong})
behaves as a constant.

\begin{figure}[t!]
\begin{center}
\subfigure[\ phantom charges on $S^3$]{\label{fig:ghostsphere}
\begin{psfrags}
\psfrag{p}[B][B][1][0]{$+$}
\psfrag{m}[B][B][1][0]{$-$}
\psfrag{a}[B][B][1][0]{$\De\a$}
\includegraphics[width=40mm]{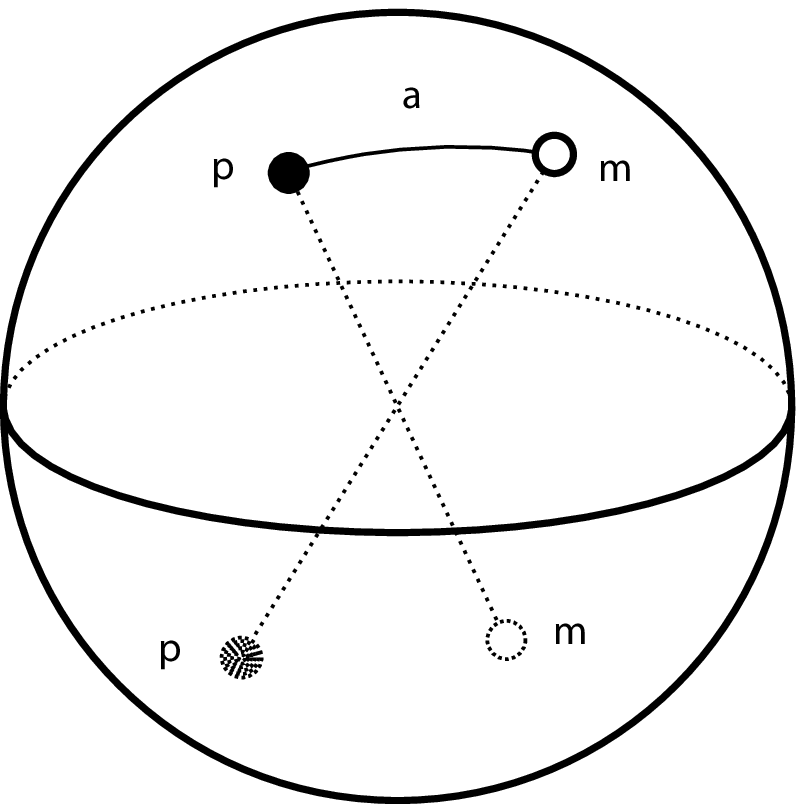}
\end{psfrags}
}
\hspace{1in}
\subfigure[\ phantom charges on $\AdS$]{\label{fig:ghosthyp}
\begin{psfrags}
\psfrag{p}[B][B][1][0]{$+$}
\psfrag{m}[B][B][1][0]{$-$}
\psfrag{b}[B][B][1][0]{$\b$}
\psfrag{q}[B][B][1][0]{$\b-i\pi$}
\includegraphics[width=45mm]{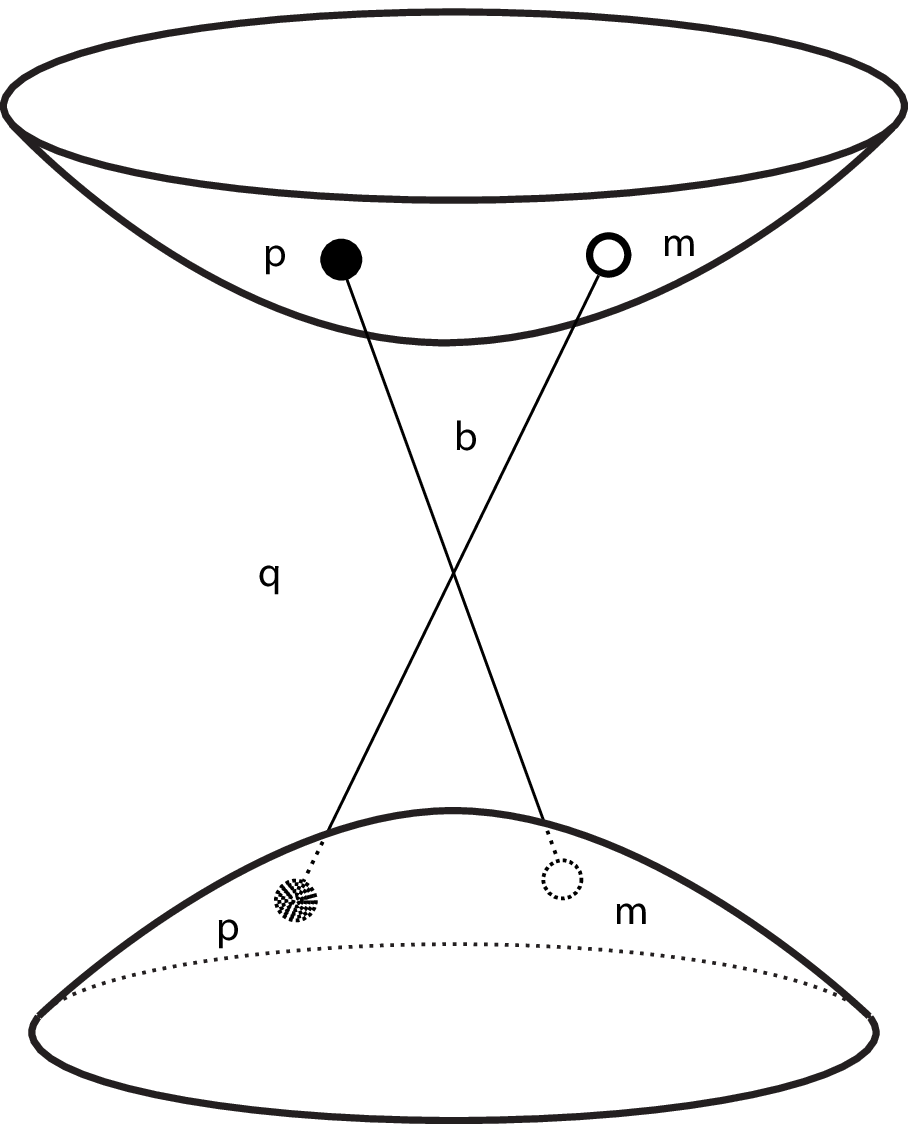}
\end{psfrags}
}
\end{center}
\caption{The naive solution to Laplace's equation on the Euclidean cylinder, Eq.~(\ref{eq:euclideanwrong}),
represents the potential in the presence of additional phantom charges at diametrically opposite points on the sphere, Figure~\ref{fig:ghostsphere}.  After analytic continuation back to Minkowski signature, the phantom charges map to another copy of AdS, Figure~\ref{fig:ghosthyp}, corresponding to phantom initial state particles.
}
\end{figure}

The problem with this potential is that it has the wrong boundary
conditions. This is easiest to understand by analytically
continuing to Euclidean space. Defining $\beta = i \alpha$, the metric
becomes
\begin{equation}
d s^2 = d \tau^2 + d \alpha^2 + \sin^2(\alpha) d
\Omega_2^2= d \tau^2 + d \Omega_3^2,
\end{equation}
which describes a Euclidean cylinder $\mathbb{R} \times S^3$.
The Wilson lines are now static charges at points on a three-sphere.
The general homogeneous solution
to Laplace's equation on $\mathbb{R} \times S_3$ is the analytic continuation of Eq.~(\ref{wrong}),
\begin{equation}
\label{eq:euclideanwrong}
  A_{\tau} (\alpha) = C_1 + C_2 \cot \alpha.
\end{equation}
Since $\cot \alpha$ has a pole at both $\alpha = 0$ and $\alpha = \pi$, Eq.~(\ref{eq:euclideanwrong}) actually describes a configuration with  two charges:
a $(+)$ charge at the north pole and
a phantom $(-)$ charge at the south pole. If we now consider two physical charges separated by
an angle $\Delta \alpha$ on the sphere, we obtain a potential for not just
these two charges, but also for two additional phantom charges, which
is incorrect. This is shown graphically in Figure~\ref{fig:ghostsphere}. The AdS version is shown in Figure~\ref{fig:ghosthyp} and discussed more below.

A nice way to get the correct solution to Laplace's equation on the Euclidean
cylinder is to add and subtract a constant charge density.  On the Euclidean
cylinder, a point charge should correspond to the source current $J_\tau =
\delta^3(x)$. Instead, we take $J_{\tau}(x) = \delta^3 (x) - \frac{1}{2
  \pi^2}$ which has a point charge at $x=0$ but is neutral overall.  If we
linearly combine such charge densities to construct an overall neutral
collection of point charges, the constant parts of the charge density will
exactly cancel, but the phantom charges will be absent. The solution to
Laplace's equation on the Euclidean cylinder with this current is
\begin{equation}
\label{eq:euclideanpotential}
  A_{\tau}^{\text{Eucl.}} (\alpha) = \frac{1}{4 \pi^2} (\pi-\alpha) \cot \alpha +
  \mathrm{constant}
\end{equation}
where the constant is an overall energy which is not yet fixed.  This same
result was computed earlier in Ref.~\cite{Belitsky:2003ys} by computing the
transition amplitude by summing over classical paths, and performing an infinite
sum of SU(2) characters.

The quantity $A_{\tau}(\alpha)$ in \eq{euclideanpotential} is the scalar
potential on the sphere due to one charge, assuming an overall neutral
distribution. To calculate the total energy for two charges $q_1=-q_2$ separated
by a distance $\alpha_{1 2}$, we can compute
\begin{equation}
E_{\mathrm{pair}}^{\text{Eucl.}}(\alpha_{12}) = \frac{1}{2} \int_{\Omega_3} (
\vec{E}_1 + \vec{E}_2)^2 ,
\end{equation}
where $\vec{E}_{1,2} = \vec\nabla (q A_{\tau})_{1,2}$ is the electric field due
to each charge. Integrating by parts, using the equations of motion, and throwing
away the infinite self-energy of each charge, this is just
\begin{equation}
  E_{\mathrm{pair}}^{\text{Eucl.}}(\alpha_{12}) = q_1q_2 A_{\tau} (\alpha_{12})+ \text{constant}.
\end{equation}
Already, the reader may recognize $E_\mathrm{pair}^{\text{Eucl.}}(\a)$ as the
$\alpha$-dependent part of the cusp anomalous dimension in Euclidean space.

The equivalent of this calculation in AdS gives
\begin{equation} \label{eq:2Jenergy}
  E_{\mathrm{pair}} (\beta_{12}) = \frac{q_1q_2}{4 \pi^2} \Big[( \pi + i \beta_{12}) \coth \beta_{12} + C \Big].
\end{equation}
Here $C$ is an undetermined constant setting the zero of energy, which must be
fixed by other considerations.  When both Wilson lines are outgoing, as in the
process $e^+e^- \to q\bar{q}$, $\beta_{12}$ is real and positive.  At small
$\beta$, the charges become closer than the curvature scale, and the energy
reduces to
\begin{equation}
E_\mathrm{pair}(\beta_{12}) \to \frac{q_1 q_2}{4\pi}\frac{1}{\beta_{12}} .
\end{equation}
This is the correct behavior of the energy of two charges as a function of
geodesic separation. Unfortunately, setting $\beta=0$ is singular, so this limit
cannot be used to determine the constant $C$.

When one Wilson line is outgoing and one is incoming, as in DIS, there is a smooth limit to zero separation which can be used to fix $C$.
In this configuration, the quantity $\gamma = \beta - i \pi$ is real. Expressing the energy
in terms of $\gamma_{12}$, we obtain
\begin{equation}
\label{eq:DISenergy}
E_{\mathrm{pair}} (\gamma_{12}) = i\frac{q_1q_2}{4 \pi^2} (\gamma_{12}
\coth \gamma_{12} -i C).
\end{equation}
In this case, the limit $\gamma_{12} = 0$ is physical: it corresponds to our two
Wilson lines reducing to a single straight line going from $t = - \infty$ to $t
= + \infty$ through the origin. This contour has no cusp and is in fact a
conserved current (occurring in the Isgur-Wise function~\cite{Manohar:2000dt}),
so its anomalous dimension must vanish. This determines the boundary condition
$E_\mathrm{pair}(\gamma_{12}=0) = 0$, which sets $C = - i$.

In summary, restoring the color factors, charges, and coupling constant for QCD,
and summing over pairs of charges to compute the total energy, we have found
\begin{equation}
  \label{eq:Etotal}
  E_{\mathrm{tot}} = \frac{i\alpha_s}{\pi} \sum_{i<j} \bm{T}_i \cdot
  \bm{T}_j \Big[ (\beta_{i j} - i \pi) \coth \beta_{i j} - 1 \Big] \,.
\end{equation}
Taking into account the factor of $i$ in going from the energy to the anomalous dimension, Eq.~(\ref{eq:matchingofdilatationandhamiltonian}),
this implies
\begin{equation}
    \Gamma = -\frac{\alpha_s}{\pi} \sum_{i<j} \bm{T}_i\.\bm{T}_j
   \left( (\beta_{i j} - i \pi) \coth \beta_{i j} - 1 \right)
\label{eq:anomo}
\end{equation}
which agrees exactly with the anomalous dimension extracted from the one-loop
calculation, Eq~\eqref{eq:1loopcuspintermsofbeta}. Thus, we have reproduced
\fact~{\bf 2} in the introduction with a simple classical calculation. Note that
with nontrivial color factors, the energy of the state in AdS corresponding to
the Wilson line operators becomes a matrix on the space of Wilson lines $\cW$
described in Eq.~(\ref{Wilsonlinewithtensor}).  This is a manifestation of the
state-operator correspondence in conformal field
theory~\cite{DiFrancesco:1997nk}.\footnote{In radial quantization, the
  operators $\cW(n_i)$ with different $t_{c_i}$ map to states in the Hilbert
  space $\mathcal{I}$ with a Hamiltonian given by Eq.~(\ref{eq:Etotal}). One
  usually considers the state-operator correspondence for local operators, which
  map to the states on $\AdS$ or $S^3$, depending on the signature.  In the
  presence of Wilson lines in the $\tau$-direction, the Hilbert space changes
  $\cH_{\AdS}\to \cH_{\AdS,\cW}$, and we can think of $\cI$ in Eq.~(\ref{I}) as
  the space of lowest-lying states in $\cH_{\AdS,\cW}$.  Interpreting the Wilson
  lines as infinitely-massive charged particles, $\cI$ is the space of lowest
  energy ``bound states" of these particles, and the anomalous dimension
  measures the finite energy differences between different bound states.
  Additional local operators would map to excitations on top of the states in
  $\mathcal{I}$.}

  Before moving on to the other \facts, it is interesting to think about the
  ``wrong'' solution, Eq.~\eqref{wrong} in AdS. On the sphere, the wrong
  solution had phantom charges on the antipoles.  The location of these phantoms
  on AdS are shifted from the location of the physical charges by
  $\beta_{ij}^{\mathrm{phant.}} = \beta_{ij} - i \pi$.  That is, the phantom is
  an initial state parton pointing in same direction as the outgoing one. So,
  for example, if we were trying to calculate $e^+ e^- \rightarrow \overline{q}
  q$, the wrong solution would have corresponded to forward (non)scattering in
  $\bar{q} q \rightarrow \bar{q} q$, depicted in Figure~\ref{fig:ghosthyp}. In
  contrast to the $e^+ e^- \rightarrow \overline{q} q$, this process has a
  smooth limit in which the $S$ matrix is just $\mathbbm{1}$.

That there is a smooth limit $\gamma\to 0$ with one incoming and one outgoing Wilson line
but not $\beta \to 0$, with two outgoing or two incoming Wilson lines is closely related to
\fact~{\bf 3} from the introduction, concerning the complexity of the anomalous dimension. The anomalous dimension is real in DIS,
since there is no obstruction to flattening the cusp. In the $e^+e^-\to $ dijets case, one
cannot remove the cusp for any geodesic separation -- the anomalous dimension has an $i\pi$ for any $\beta$.
For a single log, the $i \pi$ in the anomalous dimension can be seen to come from $\ln
\left( - \frac{\mu^2}{n_i \cdot n_j} \right)$, whose real part is the same as
$\ln \left( \frac{\mu^2}{n_i \cdot n_j} \right)$. For a double log,
\begin{equation}
  \mathrm{Re} \left[  \ln^2 \left( - \frac{\mu^2}{n_i \cdot n_j} \right) \right]
  = \mathrm{Re} \left[  \ln^2 \left( \frac{\mu^2}{n_i \cdot n_j} \right) \right]
  - \pi^2 \,.
\end{equation}
These $\pi^2$ terms get exponentiated leading to large factors of $e^{-\alpha_s
  \pi^2}$ in cross sections. In fact, this factor is a significant part of the
large $\sim 2-3$ $K$-factors in Higgs or Drell-Yan production at the
LHC~\cite{Parisi:1979xd, Sterman:1986aj, Magnea:1990zb,
  Eynck:2003fn,Ahrens:2008qu}.  Roughly, $\sigma_{\mathrm{NLO}} =
\sigma_{\mathrm{LO}} \exp \left( \gamma_{\mathrm{cusp}} (\alpha_s) C_A \pi^2
\right) \sim 3\sigma_{\mathrm{LO}}$, where the $C_A = 3$ factor comes from these
being $g g$ initial states at the LHC, and at leading order
$\gamma_{\mathrm{cusp}} (\alpha_s) = \frac{\alpha_s}{\pi} \sim 0.04$.
The AdS picture gives us a way to visualize the situations, like DIS, where the
anomalous dimension is real and situations like Drell-Yan, where it is complex.
Note that, for most processes, the anomalous dimension is a matrix, with some
real and some imaginary parts, so this picture is not tremendously useful in
general.

Next, we observe that since the Coulomb potential does not get radiative
corrections in QED (without propagating fermions), the anomalous dimension
derived with energies in AdS is also one-loop exact. This implies that the
anomalous dimension of a configuration with two Wilson lines is 1-loop exact in
QED, which is equivalent to Abelian exponentiation which was \fact~{\bf 4}.
This same reasoning applies to the potential for $N$ charges in QED. This
classical proof of Abelian exponentiation is more intuitive than the
conventional proof~\cite{Yennie:1961ad} which dissects the relevant Feynman
diagrams through repeated use of the eikonal identity and careful consideration
of diagrammatic combinatorics.

Beyond Abelian exponentiation, there are results about non-Abelian
exponentiation~\cite{Gatheral:1983cz, Frenkel:1984pz}, including
recent generalizations to multijet Wilson lines~\cite{Gardi:2010rn,Mitov:2010rp}. Non-Abelian exponentiation is not as
constraining as Abelian exponentiation, but it does imply that only a reduced set of web diagrams contribute in
perturbation theory. There are also intriguing results on the exponentiation properties of Wilson lines for quantum gravity~\cite{White:2011yy,Akhoury:2011kq,Naculich:2011ry}. It is natural to expect that there should be a way to understand these results 
using the AdS language as well, but we leave this for future consideration.

\section{One-loop results} \label{sec:oneloop}

For the anomalous dimension at 1-loop, all that is needed is the classical
Coulomb energy between two charges, as calculated in section III. It is helpful
to see how this calculation connects directly to the field theory calculation
using propagators, which will also set up the discussion of conformal gauge and
2-loop results in the next sections. We will perform most of our calculations in
the DIS case, where the cusp angle $\gamma=\cosh^{-1}(-n_1\.n_2)$ is real, since
that simplifies many of the expressions. We will also use $\gamma$ in place of
$\beta$ in our AdS coordinates.

The classical Coulomb potential $A_{\tau} (x)$ can be calculated using the AdS
propagator via
\begin{equation}
  A_{\mu} (x) = -i \int d^4 y D_{\mu \nu} (x, y) J^{\nu} (y)
\end{equation}
with the current $J^{\nu} (y)$ of a static charge. Putting the charge at the
origin we set $J^{\tau} (y) = \delta^3 (y)$ and $\vec{J} (y) = \vec 0$, and we
have
\begin{equation}
  A_{\tau} (\tau,\g) = -i \int_{- \infty}^{\infty} d \tau' D_{\tau \tau}
  (\tau,\g; \tau',0).
\end{equation}
In Minkowski space $\mathbb{R}^{1, 3}$ the position space Feynman propagator
is
\begin{equation}
  D_{\mu \nu}^F (x, y) =  \frac{1}{4 \pi^2} \frac{g^{\mu \nu}}{(x - y)^2}.
\end{equation}
Projecting onto the $\tau$ direction, this becomes
\begin{equation}
  D_{\tau \tau}^F (x, y) = D_{\mu \nu}^F (x, y) \frac{\partial
  x^{\mu}}{\partial \tau} \frac{\partial y^{\nu}}{\partial \tau}
=  \frac{1}{4\pi^2} \frac{x \cdot y}{(x - y)^2}.
\end{equation}
Finally, taking $x = e^\tau(-\cosh \gamma, -\sinh \gamma, 0, 0)$ and $y = e^{\tau'} (1, 0, 0,
0)$, we find
\begin{equation}
  D_{\tau \tau}^F = -\frac{1}{8 \pi^2} \frac{\cosh \gamma}{\cosh (\tau-\tau') + \cosh
  \gamma}, \label{eq:Dtautau}
\end{equation}
so that
\begin{equation}
  A_{\tau} (\tau,\gamma) = -i \int_{- \infty}^{\infty} d \tau' D_{\tau \tau}^F =
  \frac{i}{4 \pi^2} \gamma \coth \gamma \label{eq:Atau}.
\end{equation}
Which is the same result we found in the previous section
Eq.~(\ref{eq:DISenergy}), up to an undetermined additive constant. We can again
fix the constant by specifying which energy we mean by matching to the case of a
conserved current with $\gamma = 0$, to get $\gamma\coth \gamma-1$.

The propagator calculation lets us connect the energy directly to the
anomalous dimension. Consider the 1-loop calculation with the gluon exchanged
between the two Wilson lines. In position space, the integral is
\begin{equation}
  I = \frac{g^2}{(4 \pi)^2} \int^0_{\infty} d s \int_0^{\infty} d t \frac{
  n_1 \cdot n_2}{(s n_1 - t n_2)^2}
\end{equation}
Pulling out the overall scale and changing to radial coordinates via $t = s
e^{\tau}$ this becomes
\begin{equation}
  I = \frac{g^2}{2(4 \pi)^2} \int_0^{\infty} \frac{d s}{s} \int_{-
  \infty}^{\infty} d \tau \frac{\cosh \gamma_{}}{\cosh \tau + \cosh \gamma}
\end{equation}
where we have used $\cosh \gamma = -n_1 \cdot n_2$ as usual. The first integral
is scaleless, resulting from the fact that our configuration of Wilson lines is
rescaling-invariant.  In general, one must break rescaling invariance with UV
and IR regulators, and carefully extract the coefficients of the UV divergences
to compute the anomalous dimension.  However at 1-loop, we can be more cavalier.
With regulators in place $\int \frac{ds}{s}$ will become
$\log\frac{\L_\mathrm{UV}}{\L_\mathrm{IR}}$, so that the 1-loop anomalous
dimension is simply the coefficient of this scaleless integral. This is
precisely Eq.~\eqref{eq:Etotal}, the energy of the charges in AdS, as expected.
In the field theory calculation the $-1$ factor that appears in the energy is
correctly reproduced by the self-energy graphs for $n_i^2\ne 0$.  This
calculation makes the connection transparent at the level of the integrals.


\section{Lightlike limit}
\label{sec:lightlikelimit}
In this section, we consider the lightlike limit $n_i^2\to 0$ which was the subject of \fact~{\bf{5}}. In this limit the
static sources on the AdS space (corresponding to the Wilson lines) move towards
the boundary of AdS, $\gamma_{i j} \rightarrow \infty$ or $\beta_{ij}
\rightarrow \infty$. The anomalous dimension becomes linear in the cusp angles,
which diverge as $\beta_{ij}\to \infty$ or $\gamma_{ij}\to\infty$.
Equivalently, the imaginary part of the energy becomes linear in the geodesic
distance between the charges, while the real part goes to a constant (see
Eq.~\eqref{eq:DISenergy}). This is a qualitatively very different behavior from
flat space, where the energy vanishes as the inverse of the distance.

If we try to actually set $n_i^2 = 0$, the cusp angle is infinite and the energy
is formally infinite, indicating new unregulated singularities.  The linearly
diverging cusp angle is connected to the appearance of additional collinear
divergences that appear in both the ultraviolet and infrared in the anomalous
dimension computation. Effectively in the computation of matrix elements of
$\cW$ we are forced to introduce a small dimensionful IR regulator $\Lambda$.
With UV divergences regulated in dimensional regularization the dimensions are
compensated by $\mu$, so $1/|n| \to \Lambda/\mu$, giving the form in
Eq.~(\ref{eq:onelooplight}).  Introducing $\Lambda/\mu$ is equivalent to moving
light-like charges away from the boundary of AdS, so that we can still sensibly
talk about the geodesic distance between charges. The dependence on the IR
regulator cancels out in physical cross section computations. For example,
eikonal scattering involves a square of $\cW$ matrix elements which yields the
soft function $S(k)$ in \eq{S}.  In this case the IR divergences cancel between
virtual and real emission diagrams and $\Lambda$ is replaced by the observed
momenta of particles, $\Lambda\to k$.

\begin{figure}
\begin{center}
$
\begin{array}{ccc}
 \includegraphics[width=0.3\textwidth]{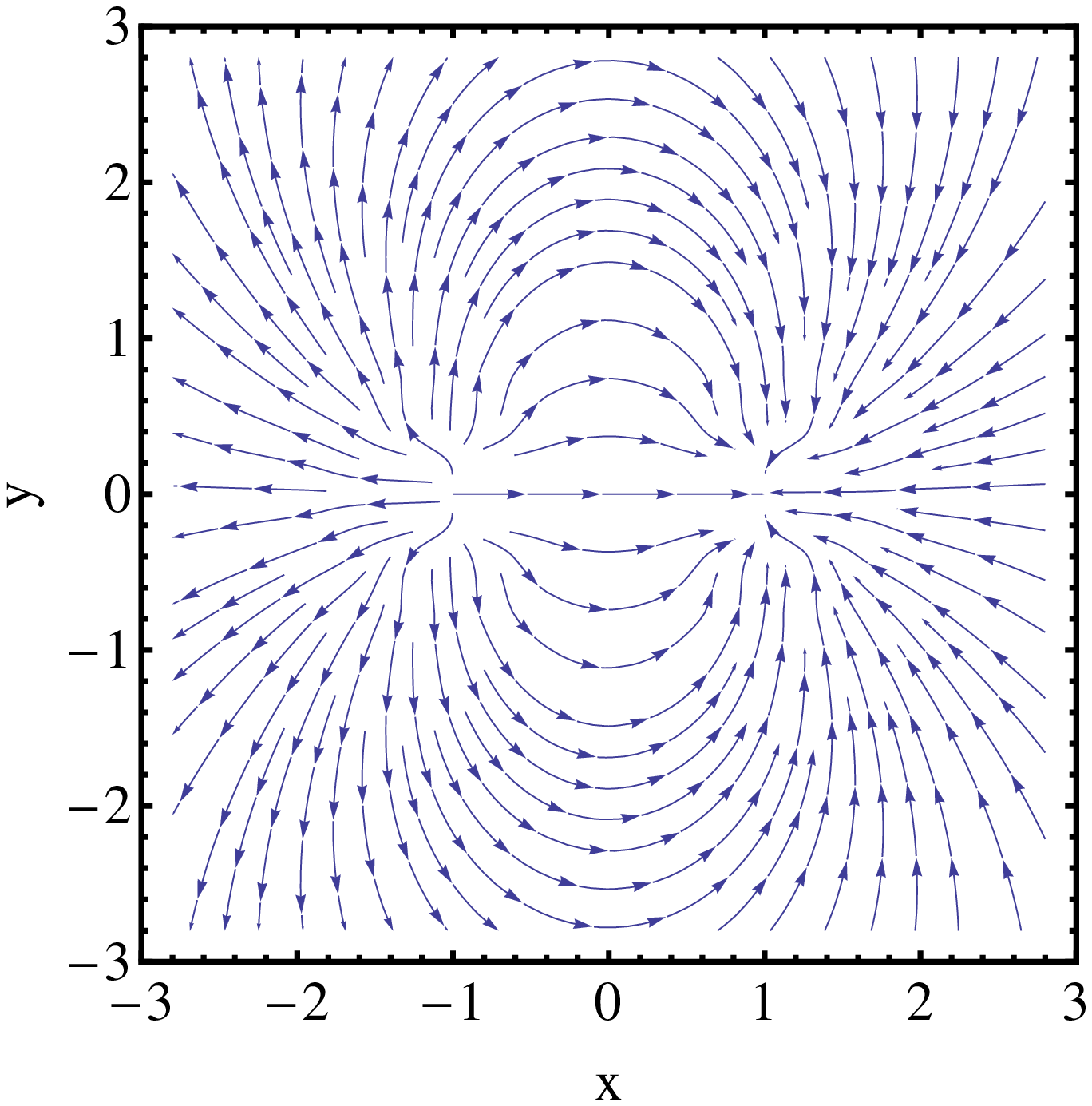} &
 \includegraphics[width=0.3\textwidth]{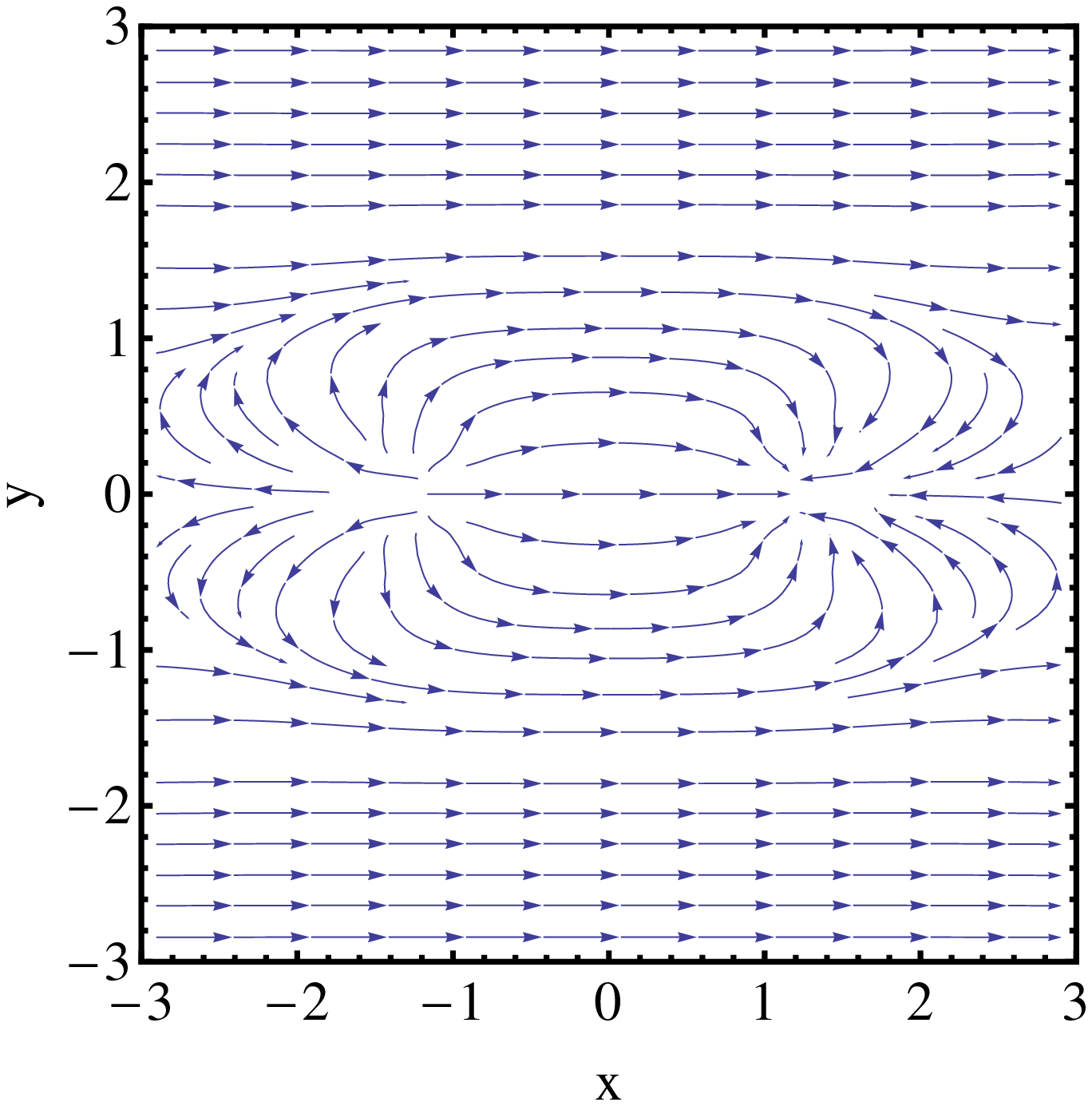} &
  \includegraphics[width=0.3 \textwidth]{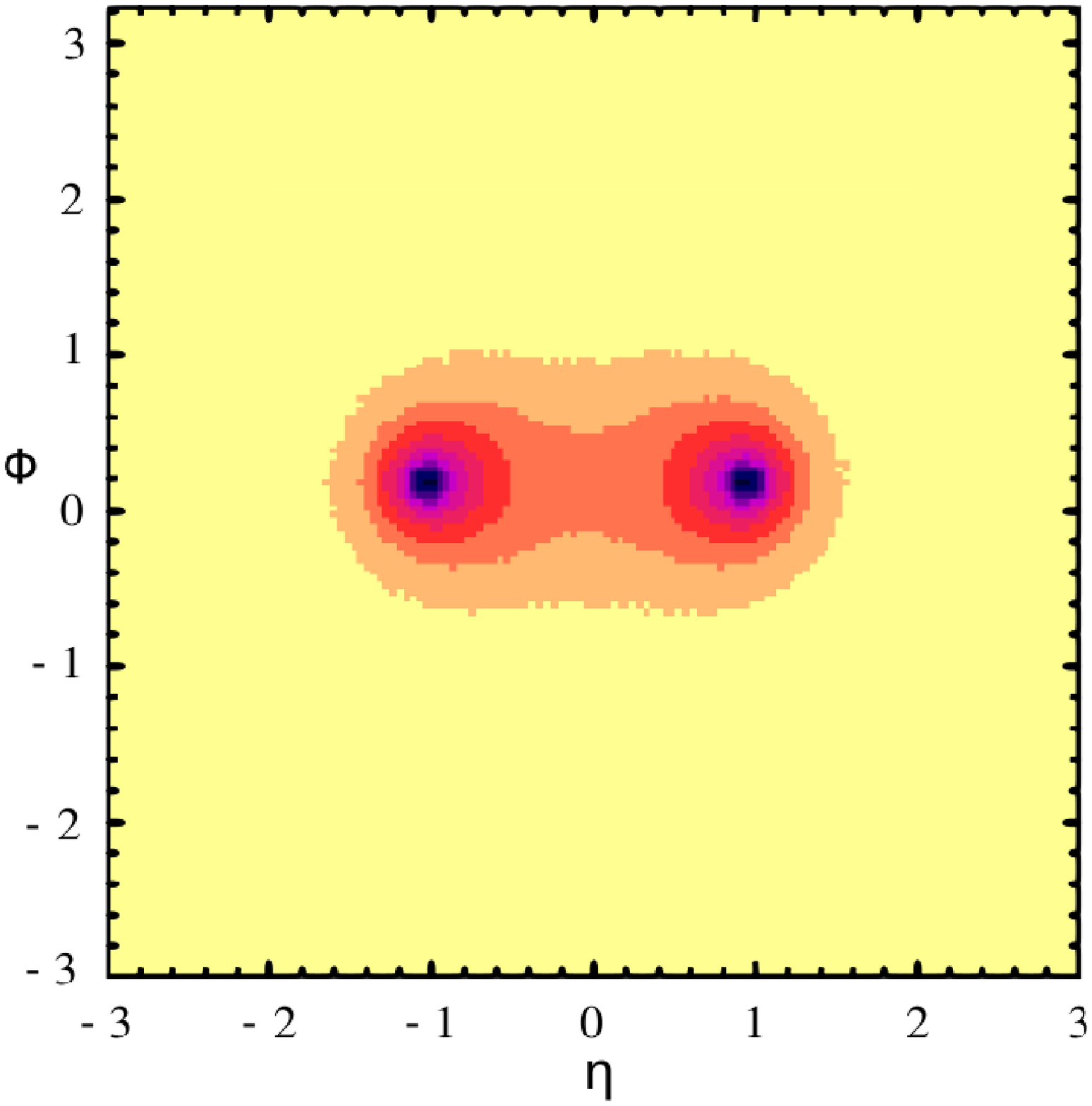}
\end{array}
$
\end{center}
\caption{On the left is the electric field lines for two charges in flat space. The middle shows the imaginary part of the electric
field for two charges in AdS, after projecting to rectangular coordinates with $x=\beta \sin\theta$ and $y=\beta \cos\theta$.
The right (from~\cite{Gallicchio:2010sw}), shows the distribution of radiation from a color singlet scalar decaying to two jets at the LHC. The axes in this case
are psuedorapidity and azimuthal angle, and the contours correspond to factors of two in the accumulated
energy distribution.
The rightmost plot is included to remind the reader that a color dipole radiates between the color charges, which roughly corresponds
to the region where the energy density has support in  the AdS picture. The sharp drop-off of the radiation pattern in the
effect of color coherence. In a qualitative sense only, this corresponds to the exponential decay of the radiation away from the dipole axis in the
AdS picture.
 \label{fig:sudconf}}
\end{figure}

To secure a clearer physical picture of what this growing imaginary energy
means, consider the case of two outgoing lightlike partons, such as in dijet
production. In Figure~\ref{fig:sudconf}, we contrast the electric field of a
normal pair of charges, in flat space, with the imaginary part of the electric
field for two charges in AdS. As the total energy grows with separation, the
electric field approaches a constant between the charges. Notice that in the
presence of two opposite Abelian charges, $\Im(E)$ is always negative, as should
be the case for the energy of an unstable state.  Back in Minkowski space, this
corresponds to a roughly constant density of radiation between the two charges.
The third panel of Figure~\ref{fig:sudconf}, shows this behavior in a Monte
Carlo simulation~\cite{Sjostrand:2006za}.  To generate this distribution, a 200
GeV dijet event produced at 7 TeV center of mass energy at the LHC was
simulated.  The figure shows the accumulated energy distribution.  Note that the
radiation is concentrated between the two charges, and suppressed
away from the dipole axis, just as the energy distribution is in AdS.

Linear growth of energy with separation is normally an indication of confining
behavior. In this case, since the energy is imaginary, it is not confinement in
the usual sense, but can still be interpreted as a type of confinement.
In a sense, this linear growth of the (imaginary) energy with separation is
related to the fact that high energy quarks always appear with an accompanying
jet, whose dynamics are described with Sudakov factors. Although this ``Sudakov
confinement" of quarks inside jets has little in common with confinement in QCD,
it is not an unreasonable phrase for the linearly growing energy in AdS.

To be specific, consider the case of one incoming and one outgoing Wilson line,
as in deep-inelastic scattering (DIS). In DIS, an electron is scattered off of a
proton, with Bjorken $x$ defined in terms of the momentum transfer $q^\mu$ and
the proton momentum $P^\mu$ as $x = -q^2/2 P\cdot q$. In the $x\to1$ limit of
DIS at large $Q^2=-q^2$, the outgoing radiation becomes jet-like, with its mass
$m_J^2 = Q^2(1-x)/x$ getting small. In this limit, the Wilson line description
applies and important physics is encoded in the anomalous dimension.  We saw
that when Wilson lines becomes lightlike, the cusp angle $\gamma_{12}\to \infty$
and the energy in AdS becomes
\begin{equation}
  E_{\mathrm{tot}} \to i \frac{\alpha_s}{\pi} \gamma_{12}
\end{equation}
which is an imaginary energy.

An imaginary energy is indicative of non-unitary time evolution.
Since time in AdS is scale in Minkowski space, this is non-unitary evolution
as the scale is changed. One way to think of the violation of unitary is
going from a simple state, with say a single quark, to a state with a quark and a gluon.
The probability for this to happen is given by the Alterelli-Parisi splitting kernels.
The quark-gluon splitting kernel is
\begin{equation}
 P_{q\to q g}(z,p_T) = \frac{\alpha_s}{\pi} \frac{1+z^2}{1-z}\frac{1}{p_T}
\end{equation}
where $z$ is the fraction of energy in the quark, which must be close to 1 (the Wilson line picture only applies in the $x\to 1$ limit)
and $p_T$ is the transverse momentum of the gluon with respect to the quark direction.
The integral over this splitting function is infrared divergent. Typically, one imagines
the quark starts off at a scale $\mu\sim Q$ characteristic of the hard scattering; then
it evolves down to a scale $\mu\sim p_T$~\cite{Bauer:2006mk,Bauer:2006qp}. The probability of not having emitted
a gluon between these scales is
\begin{equation}
 \Sigma(Q,p_T) = \exp \left(- \int_{p_T}^{Q} \rd \mu \int_{Q/\mu}^1
   \rd z\  P(1-z,\mu)\right)
= \exp\left(-\frac{\alpha_s}{\pi}\ln^2\frac{Q}{p_T}+\ldots \right).
\end{equation}
This quantity is known as a Sudakov factor. It is a no-branching probability,
and used as a classical probability in Monte Carlo event generators, which treat
the parton as showering off gluons through a Markov process, see for
example~\cite{Sjostrand:2006za}. The rate of emission is exactly proportional to
the coefficient of linear growth of the energy in AdS, a quantity known as the
{\it cusp anomalous dimension}.

So we see that the non-unitary evolution corresponds to the probability that an
off-shell quark decays into a gluon and a slightly less off-shell quark.  The
constant energy density in AdS at large cusp angle corresponds to a constant
probability for emission per unit scale.  If we evaluate the Sudakov factor at
$p_T=0$, we see that there is zero probability for a quark to evolve forever
without emitting a gluon. In physical situations, there is always a finite IR
cutoff, such as the hadronization scale $\Lambda_{\text{QCD}}$ or a resolution
scale for the jet. Nevertheless, at late times in AdS, or small momentum scales
in Minkowski space, the original state approaches zero amplitude. Thus, there is
no probability that we will find a free quark at asymptotically late times when
produced at a finite scale $Q$. In other words, there are no free quarks which
are not confined within jets. In fact, the characteristic size of jet is
precisely determined by the cusp anomalous dimension in the Sudakov factor.

It would be interesting to further explore the analogy between confinement and
the linear growth of (imaginary) energy with distance. For example, one might
argue that the energy can grow at most linearly with
separation~\cite{Sundrum:1997qt}, as expected from the string flux tube picture.
The analogy would connect this directly to Sudakov confinement and the linear
dependence of the anomalous dimensions on $\log\mu$ for various fields in
scattering processes. Then, perhaps, by reversing the logic, the Sudakov factors
could give insight to confining gauge theories from the study of jets.

\section{Conformal gauges} \label{sec:conformalgauge}

We have seen that radial quantization is a useful picture for understanding the
one-loop cusp anomalous dimension.  In $\R\x\AdS$, a cusp maps to a collection
of charges, and $\G^\mathrm{1-loop}_\mathrm{cusp}$ has an interpretation in
terms of static potentials.  It makes sense that for higher-loop computations,
we should keep the separation between the $\tau$ direction and the AdS directions
manifest.  However, this is not done in typical calculations.  The reason is
that although the action $S_\mathrm{YM}$ is conformally invariant, the {\it
  gauge-fixing terms} are not.  For example, in Feynman gauge, we have
\begin{align}
S =& S_\mathrm{YM}+S_\mathrm{g.f.}, \nn\\
S_\mathrm{g.f.} =& \int \rd^4 x \sqrt{-g}\frac 1 2 (\nabla^\mu A_\mu)^2+S_\mathrm{ghost}.
\end{align}
Under a conformal rescaling $g_{\mu\nu}=e^{2\w}g'_{\mu\nu}$, the covariant
derivative transforms nontrivially,
\begin{equation}
\label{eq:covderivtransformation}
\sqrt{-g}(\nabla^\mu A_\mu)^2 
  = \sqrt{-g'}(\nabla'^\mu A_\mu + 2\ptl^\mu \w A_\mu)^2.
\end{equation}
Consequently, a gauge that appears natural in one conformal frame may look
somewhat unnatural in another.  For instance, a useful feature of Feynman gauge
in flat Minkowski space is that the propagator does not mix different
polarizations between different points: $\e^\mu \Pi_{\mu\nu}\propto\e_\nu$.
However from the point of view of radial coordinates $(\tau,\b,\th,\f)$, the
Feynman gauge propagator transports polarizations from one point to another in a
nontrivial way. More precisely, transport via the propagator is not proportional
to parallel transport in the $\R\x\AdS$ metric. This is clear from the
transformation law~(\ref{eq:covderivtransformation}): plugging in
$g_{\R^{1,3}}=e^{2\tau}g_{\R\x\AdS}$, we see that $S_\mathrm{g.f.}$ includes
kinetic mixing between $A_\tau$ and the spatial components $A_i$. Throughout
this section, we will use $i=1,2,3$ to denote the directions in AdS.

It is informative to look at the explicit form of the mixing. To decompose the
propagator, first observe that the $\tau$ component of the propagator comes
from projections onto $\frac{\partial x^{\mu}}{\partial \tau} = x^{\mu}$. So
that $D_{\tau \nu} = x^{\mu} D_{\mu \nu}$. Thus we can decompose the polarization
$g^{\mu\nu}$ in the numerator of the Feynman propagator as
\begin{align}  \label{gmndec}
g^{\mu \nu} =
& \Big[
g^{\mu \nu} - \hat{x}^{\mu} \hat{x}^{\nu} - \hat{y}^{\mu} \hat{y}^{\nu}  + ( \hat{x} \cdot \hat{y}) \hat{x}^{\mu} \hat{y}^{\nu}
\Big]
+
 \Big[
( \hat{x} \cdot \hat{y}) \hat{x}^{\mu} \hat{y}^{\nu}
\Big]
\nn \\
&
 +\Big[
\hat{x}^{\mu} \hat{x}^{\nu} + \hat{y}^{\mu} \hat{y}^{\nu}
-2 ( \hat{x}  \cdot \hat{y}) \hat{x}^{\mu} \hat{y}^{\nu}
 \Big]
 \,,
\end{align}
where $\hat{x} ={x^{\mu}}/{|x|}$ and $\hat{y} = {y^{\mu}}/{|y|}$. The
first term in brackets vanishes when contracted with $x^{\mu}$ or $y^{\nu}$, so
it represents the spatial components of the Feynman gauge propagator from the point of view of
AdS, i.e. $D_{ij}$. This can also be seen by writing it as
\begin{equation} \label{gdec}
  g^{\mu \nu} - \hat{x}^{\mu} \hat{x}^{\nu} - \hat{y}^{\mu} \hat{y}^{\nu} + (
  \hat{x} \cdot \hat{y}) \hat{x}^{\mu} \hat{y}^{\nu} = |x||y|
  \frac{\partial}{\partial x_{\mu}} \frac{\partial}{\partial y_{\nu}}
  \left( \frac{x \cdot y}{|x| |y|} \right)
  \,,
\end{equation}
which vanishes when contracted with $x_{\mu}$ or $y_{\nu}$ since the dilation
operator $\mathcal{D} = x^{\mu} \partial_{\mu}$ automatically annihilates any
scale-invariant function. The second term in Eq.~(\ref{gmndec}) is $D_{\tau
  \tau}$, as in Eq.~\eqref{eq:Dtautau}.  The third term represents the nonzero
$D_{\tau i}$ and $D_{i\tau}$ mixing present in Feynman gauge.

Since the Wilson lines only source $A_{\tau}$, at one-loop only the $D_{\tau
  \tau}$ component of the propagator contributes. If we try to use the same
propagator in higher-loop computations, even though the Wilson lines only source
$A_{\tau}$, due to the $D_{\tau i}$ mixing terms, there will be interactions
involving $(A_i)^3$ vertices which make the calculations complicated. However,
we have seen that anomalous dimensions of Wilson lines are most naturally
thought about from the point of view of $\R\x\AdS$.  In a gauge more suited to
this space there should be no mixing, and $A_\tau$ can be treated as a charged
scalar. Since there is no scalar cubic or quartic vertex $(A_\tau)^{3,4}$ in
Yang-Mills theory, the calculation will be significantly simpler.  Vertices
$(A_i)^2 A_\tau$ remain active when we consider Wilson line operators in this
gauge and enter for the leading vacuum polarization effects.

We will refer to gauges with the property that $A_\tau$ and $A_i$ do not mix as
{\it conformal gauges}, to emphasize the fact that they are most natural in a
different conformal frame from the usual one.  This non-mixing is not an overly
restrictive condition and there are many gauges that satisfy it (for example,
the condition remains true after any $\tau$-independent gauge transformation).
Perhaps the simplest example of a conformal gauge is temporal gauge in
$\R\x\AdS$, or equivalently ``radial gauge" in $\R^{1,3}$, in which $A_\tau=
x^\mu A_\mu=0$ and radial Wilson lines are actually trivial.  We will briefly
discuss this gauge in Section~\ref{sec:radialgauge}.  In
\sec{conformalgaugederivation} we will focus on a less singular example of
conformal gauge, which corresponds to a quantum average over different
gauge-conditions, as in $R_\xi$ gauges.

\subsection{Derivation of Conformal Gauge in $d$-dimensions.}
\label{sec:conformalgaugederivation}

To arrive at a conformal gauge, perhaps the most familiar strategy would be
to study gauge-fixing terms in $\R\x\AdS$, and then invert the
kinetic terms to form the corresponding propagators.  This might be an
interesting exercise, but it would be needlessly complicated for our purposes.
Instead, we will adopt the more pragmatic procedure of directly
gauge-transforming the Feynman-gauge propagator and solving for the
transformation function that gives the desired properties.  This then implicitly
specifies BRST exact gauge-fixing terms, including a ghost action.

Our goal is to derive a gauge that has no $\tau$-$i$ mixing in $d$-dimensions
that is suitable for use in dimensional regularization. Our procedure is simple,
and powerful enough to handle this even though Yang-Mills theory is only
classically conformally invariant when $d=4$.  Instead of sorting out details of
conformal anomalies at order $\e$, we will keep the $d$-dimensional Minkowski
metric and flat coordinates $x_\mu$, and simply search for a gauge that respects
the foliation of $\R^{1,d-1}$ into $\R\x\AdS_{d-1}$.  Precisely when $d=4$, our
propagator will have an interpretation as the propagator in a gauge theory on
$\R\x\AdS_3$.  However when $\e$ is nonzero, it will simply be a useful tool
that enables computations to be performed without encountering mixing terms
$D_{\tau i}$ or $D_{i\tau}$ in $d$-dimensions.

We begin with the position-space Feynman propagator in $d$ dimensions
\begin{equation}
D^{F}_{\mu\nu}(x,y)=- g_{\mu\nu}\, \frac{\ka_d}{ [-(x-y)^2+i\epsilon]^{d/2-1} } \label{DFd}
\end{equation}
where $\ka_d=\frac{\G(d/2-1)}{4\pi^{d/2}}$ is a constant.  Since the propagator
depends only on quadratic terms in the action, it is sufficient for our
discussion here to consider an Abelian theory.  In a non-Abelian theory, the
propagator should also include a factor of the identity $\delta^{ab}$ in color
space, and we have the additional Feynman rules involving ghosts which we
discuss in Appendix~\ref{app:ghost}.

We will consider a class of propagators given by
\begin{equation} \label{eq:Dmunu}
D_{\mu\nu}(x,y) = D^F_{\mu\nu}(x,y)+\frac{\ptl}{\ptl y^{\nu}}\L_\mu(y,x)+\frac{\ptl}{\ptl x^{\mu}} \L_{\nu}(x,y),
\end{equation}
where $\L_\mu(y,x)$ is a one-form at $x$, depending on both $x$ and $y$.  Notice
that $D_{\mu\nu}$ is still an inverse for the kinetic term in the space
of gauge equivalence-classes, though it differs from $D^F_{\mu\nu}$ along
gauge-orbits.  Indeed, suppose $J^\mu$ is a conserved current, and consider the
associated vector potential
\begin{align}
A_\mu^\conf(x) &\equiv -i\int \rd y\,D_{\mu\nu}(x,y)J^\nu(y)
  \nn \\
&= -i\int \rd y\,D^F_{\mu\nu}(x,y)J^\nu(y)+\frac{\ptl}{\ptl x^\mu}\p{-i\int \rd
  y\,\L_\nu(x,y)J^\nu(y)}
  \nn \\
&=  A^F_\mu(x) +\ptl_\mu \p{-i\int \rd y\,\L_\nu(x,y)J^\nu(y)}
  \,, \label{eq:confgt}
\end{align}
where we have integrated by parts and used current conservation.  Since
$A_\mu^\conf$ differs from $A^F_\mu$ only by a gauge transformation, it still
solves Maxwell's equation $\ptl_\mu F^{\mu\nu}=J^\nu$.

We would like $D_{\tau i}$ to vanish, so that the propagator does not mix time
and space directions in $\R\x\AdS$.  Recalling that $\tau$ is the generator of
scale transformations, $\ptl_\tau = x^\mu \ptl_\mu$, a vector field $A_\mu(y)$
will have no $\tau$ component if $y^\mu A_\mu(y) = 0$.  Thus our condition is
\begin{equation}
\label{eq:adsgaugecondition}
x^\mu D_{\mu\nu}(x,y)A^{\nu}(y) =0 \qquad \textrm{whenever}\qquad y_\nu A^\nu(y)=0.
\end{equation}
Likewise, for $D_{i\tau}$ to vanish we have the condition $A^\mu(x)
D_{\mu\nu}(x,y)y^\nu=0$ whenever $x_\mu A^\mu(x)=0$.  Note that
Eq.~(\ref{eq:adsgaugecondition}) is not translation invariant, it treats the
origin as a special point and yields propagators that are not simply functions
of $x-y$.\footnote{In our setup the origin is special since it is the location
  of our hard interaction and the place where cusps occur between Wilson lines.
  This explains why it is useful to consider non-translationally invariant
  gauges, even though the final physical results are gauge independent and
  translationally invariant.} The general class of conformal gauges which
satisfy these no-mixing conditions is derived in
Appendix~\ref{app:genconformal}. Here let us consider the ansatz
\begin{equation}
\L_\mu(y,x) = \ka_d\frac{x_\mu}{|x|^{d-2}}\, g(\a,\b),\qquad\textrm{where}\quad \a\equiv\frac{x\.y}{|x||y|},\ \b\equiv\frac{|y|}{|x|},
\end{equation}
and $g(\a,\b)$ is some function to be determined.
Equation~\eqref{eq:adsgaugecondition} implies
\begin{align}
 \pdr{}{\a}g(\a,\b) &= \b(2\a\b-\b^2-1)^{1-d/2} \nn\\
g(\a,\b) &= \frac{1}{4-d}\Big[ (2\a\b-\b^2-1)^{2-d/2}-f(\b)^{2-d/2} \Big],
\end{align}
where $f(\b)$ is arbitrary and its $2-d/2$ power produces the correct $d\to
4$ solution. Our ansatz becomes
\begin{equation} \label{eq:Lgt}
\L_\mu(y,x) = \frac{\ka_d}{4-d}\, \frac{x_\mu}{x^2}
  \Big\{[-(x-y)^2]^{2-d/2}-|x|^{4-d}f(\b)^{2-d/2} \Big\} \,.
\end{equation}

The conformal gauge propagator is then
\begin{align}
D_{\mu\nu}(x,y) =& - \frac{\ka_d}{[-(x-y)^2]^{d/2-1}}
  \Big(g_{\mu\nu}-\frac{x_\mu x_\nu}{x^2}-\frac{y_\mu
    y_\nu}{y^2}+\frac{2x_\mu (x\.y)y_\nu}{x^2 y^2} \Big) \nn\\
&+\ka_d\frac{x_\mu y_\nu}{x^2y^2}
  \Big\{ [-(x-y)^2]^{2-d/2}- \chi\big(|x|,|y|\big)^{4-d} \Big\}
\end{align}
where $\chi(|x|,|y|)$ is any symmetric function of $|x|$ and $|y|$ with dimensions of length. ($\chi$ has a
 a simple but unenlightening relation to $f(\beta)$).  Separating out
the $\tau$ and spatial components, as in Eq.~\eqref{gmndec}, this can be written
\begin{align} \label{eq:propangularradial}
D_{\mu\nu}(x,y) =& - \frac{\ka_d}{[-(x-y)^2]^{d/2-1}} {|x||y|}\ptl^x_\mu
\ptl^y_\nu\p{\frac{x\.y}{|x||y|}}
 \nn \\
&-\ka_d\frac{x_\mu y_\nu}{x^2y^2}\p{\frac{x\.y}{[-(x-y)^2]^{d/2-1}}-[-(x-y)^2]^{2-d/2}
  +\chi\big(|x|,|y|\big)^{4-d}}.
\end{align}
Here the first term is manifestly ``angular", involving derivatives acting on a scaleless quantity which vanish when
contracted with $x^\mu$ or $y^\nu$, as in Eq.~\eqref{gdec}. The second term is ``radial", involving projection onto the $\tau$-direction with $x_\mu$ and $y_\nu$.
The mixing terms have been gauged away, as desired.

A natural choice is to take $\chi$ to be $d$-independent. Then when $d=4$ the
last two terms of \eq{propangularradial} cancel, $\chi$ drops out, and we have a
unique 4D propagator. In fact, this propagator is identical to the Feynman
propagator in 4D, Eq.~\eqref{eq:Dtautau}, without the mixing terms. In
particular, the calculation of the Coulomb potential from a static charge in
AdS, and hence the one-loop anomalous dimension of the Wilson line, is identical
in Feynman gauge and in conformal gauge.  We see that the entire content of this
gauge fixing is to move the mixing terms in Feynman gauge into non-mixing terms starting
at order $\e$.

Equation~(\ref{eq:propangularradial}) is convenient for computations involving
non-light-like Wilson lines in $d$-dimensions or with dimensional
regularization.  To consider propagation between points on two light-like Wilson
lines we take $x^\mu = \lambda_1 n_1^\mu$ and $y^\mu = \lambda_2 n_2^\mu$, and
take the limit $n_1^2\to 0$ and $n_2^2\to 0$. Here $n_1^\mu n_2^\nu
D_{\mu\nu}(x,y)$ reduces to the same result as Feynman gauge in 4D.  In
$d$-dimensions the result in \eq{propangularradial} is not convenient because
the $\chi$ term does not scale in the same manner as the other terms in the
radial part of the propagator. We derive an alternative conformal gauge with a
good scaling limit for light-like Wilson lines in $d$-dimensions in
Appendix~\ref{app:genconformal}.


\subsection{Comparison to radial gauge}
\label{sec:radialgauge}

In conformal gauge, the scalar modes in AdS, $A_\tau$ which are produced from
the Wilson lines, have no mixing with the vector modes. This simplifies some
loop calculations, as we will demonstrate in the next section.  However, it is
natural to ask why we cannot simplify things even further by choosing temporal
gauge $A_\tau=0$ in AdS.  This condition becomes $x_\mu A^\mu(x)=0$ in Minkowski
space, and gauges satisfying it are called radial (or Fock-Schwinger) gauges.
The origin is again a special point for these gauges and the gauge boson
propagator is not translation invariant.

In radial gauge, our $N$-jet Wilson lines become trivial, and loop corrections
to the expectation value $\langle \cW\rangle$, and corresponding anomalous
dimension, seem na\"ively to vanish.  Of course this is too simplistic to be
correct. The problem is that an $N$-jet Wilson line operator $\cW$ as defined in
Eq.~(\ref{Wilsonlinewithtensor}) is only invariant by itself under gauge
transformations which vanish at infinity.  However, the transformation from,
say, Feynman gauge to radial gauge is nontrivial at infinity, so the expectation
value $\<\cW\>$ can change.  Indeed, in Ref.~\cite{Leupold:1996hx} it was shown
that the radial gauge propagator $D_{\mu\nu}$ itself carries ultraviolet
divergences.

To correctly compute the cusp anomalous dimension, we must either restrict
ourselves to gauges with appropriate behavior at spatial infinity,
or ``close off" our Wilson loop in a gauge-invariant way at some large finite
distance from the origin, without introducing additional cusps. (Or with
additional cusps whose contribution we then subtract away.)  In the latter case,
the cusp divergences are generated by a different part of the calculation. For
example, using a conventional definition of radial gauge,
Ref.~\cite{Leupold:1996hx} explicitly demonstrates that the classic one-loop
$x=0$ cusp anomalous dimension is correctly reproduced by the (smooth) ``closed
off'' part of the Wilson loop. The fact that the radial gauge propagator itself
is ultraviolet divergent plays a crucial role in this computation, since
otherwise the closed off part of the loop would not contribute to the anomalous
dimension.

To avoid having these complications, we will focus on conformal gauges that do
not have ultraviolet divergences in $D_{\mu\nu}$. This was true of our
construction in \sec{conformalgaugederivation}, where $D_{\mu\nu}$ in
Eq.~(\ref{eq:propangularradial}) is finite as $d\to 4$, and divergences occur only when interaction points approach each other.  In
the limit $(x-y)^2\to 0$, with both $x^2,y^2\ne 0$, $D_{\mu\nu}$ approaches the
usual Feynman propagator at leading order.  Consequently, power counting shows
that divergences originating near points away from the light-cone are identical
in Feynman gauge and in our conformal gauge. (It would be useful to fully
characterize the divergence and subdivergence structure of multi-loop diagrams in conformal
gauge.)  In these cases, the cusp anomalous dimension can be computed by
considering only a neighborhood of the cusp.


\section{Three Wilson Lines at Two-Loops}\label{sec:2loops}

As a concrete application of using our conformal gauge, let us compute the
two-loop contribution to the anomalous dimension of a multi-Wilson line operator
Eq.~(\ref{Wilsonlinewithtensor}) that involves all three lines.  The absence of
$\tau$-$i$ mixing makes this computation extremely simple, and elucidates the
origin of the previously mysterious pairwise structure of the result, discussed as \fact~{\bf{6}}.

\begin{figure}
\subfigure[\ nonplanar]{\label{fig:2loop-nonplanar}
\begin{fmffile}{2loopcusp-nonplanar}
\begin{fmfgraph*}(90,90)
\fmfset{arrow_len}{3mm}
\fmfsurroundn{v}{12}
\fmf{plain,tension=1.2}{v1,a1}
\fmf{plain,tension=1.8}{v4,b1}
\fmf{plain,tension=1.5}{v6,c1}
\fmf{plain}{a1,v9}
\fmf{plain}{b1,v9}
\fmf{plain}{c1,v9}
\fmffreeze
\fmfdot{v}
\fmf{phantom,tension=.9}{v2,v}
\fmf{photon,right=.2,tension=.8}{a1,v}
\fmf{photon,left=.1,tension=.4}{b1,v}
\fmf{photon,left=.15,tension=1.5,rubout=4}{c1,v}
\end{fmfgraph*}
\end{fmffile}
}
\subfigure[\ planar]{\label{fig:2loop-planar}
\begin{fmffile}{2loopcusp-planar}
\begin{fmfgraph*}(90,90)
\fmfset{arrow_len}{3mm}
\fmfsurroundn{v}{12}
\fmf{plain,tension=1.5}{v1,a1}
\fmf{plain,tension=1.3}{v4,b1}
\fmf{plain,tension=4}{b1,b2}
\fmf{plain,tension=1.5}{v6,c1}
\fmf{plain}{a1,v9}
\fmf{plain}{b2,v9}
\fmf{plain}{c1,v9}
\fmffreeze
\fmf{photon,right=.2}{a1,b1}
\fmf{photon,right=.2}{b2,c1}
\end{fmfgraph*}
\end{fmffile}}
\subfigure[\ counterterm]{\label{fig:2loop-ct}
\begin{fmffile}{2loopcusp-ct}
\begin{fmfgraph*}(90,90)
\fmfset{arrow_len}{3mm}
\fmfsurroundn{v}{12}
\fmf{plain,tension=1.3}{v1,a1}
\fmf{plain,tension=1.3}{v4,b1}
\fmf{plain}{v6,v9}
\fmf{plain}{a1,v9}
\fmf{plain}{b1,v9}
\fmfv{decor.shape=circle,decor.filled=empty,decor.size=8,label=$\x$,label.dist=0}{v9}
\fmffreeze
\fmf{photon,right=.2,tension=.8}{a1,b1}
\end{fmfgraph*}
\end{fmffile}}
\caption{\label{fig:2loopgraphs}2-loop graphs contributing to the coefficient $F(\g_{ij},\g_{jk},\g_{ki})$ of the antisymmetric color structure in $\G_\mathrm{cusp}(v_i)$ (Eq.~\ref{eq:2loopcusp}).}
\end{figure}
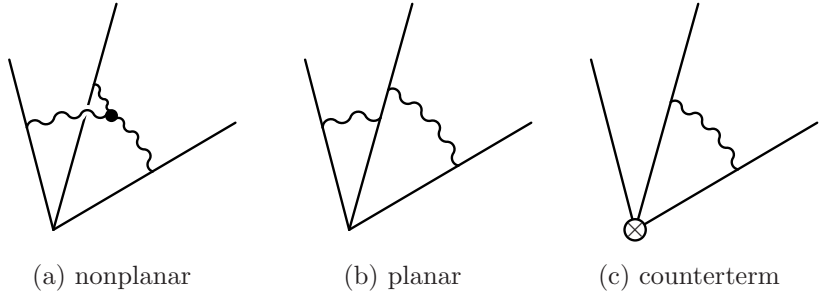

When the number $N$ of jet directions is three or more,
the anomalous dimension $\G_\mathrm{cusp}(n_i)$ can in principle depend on arbitrary combinations of the cusp angles
$\g_{ij}$.  Nontrivial combinations involving three $\g_{ij}$'s can appear first at two
loops in the coefficient $F$ of the ``maximally non-Abelian" color structure $f^{abc}\bm T_i^a \bm T_j^b \bm T_k^c$,
\begin{equation}
\label{eq:2loopcusp}
\G^\mathrm{2-loops}_\mathrm{cusp}(n_i)
= \p{\frac{\alpha_s}{\pi}}^2\p{\sum_{i<j}\bm T_i^a \bm T^a_j\,f(\g_{ij})
 +\sum_{i<j<k}if^{abc}\bm T^a_i \bm T^b_j \bm T^c_k\,F(\g_{ij},\g_{jk},\g_{ki})},
\end{equation}
due to the presence of diagrams depicted in Figure~\eqref{fig:2loopgraphs}.  In
particular the non-planar diagram~\ref{fig:2loop-nonplanar} could na\"ively
contribute a complicated function of all three cusp angles.  The expression for
this graph in Feynman gauge was discussed in the lightlike limit in
\cite{Aybat:2006wq} and \cite{Aybat:2006mz}, and analyzed numerically in
\cite{Mitov:2009sv}.  It was finally computed for general cusp-angles in
\cite{Ferroglia:2009ii} in a somewhat technical computation using Mellin-Barnes
representations.  After all this, the final result turns out to be remarkably
simple,
\begin{equation}
\label{eq:nonplanarfeyngauge}
F^\mathrm{(a)}_{\mathrm{Feyn.}}= - \frac{1}{2}(\g_{ij}\coth\g_{ij})\g_{jk}^2+\textrm{antisym.}
\end{equation}
where ``antisym." stands for signed permutations of $i,j,k$. This is a sum of
terms each of which only depends on two of the cusp angles.

It is less surprising that the planar and counterterm graphs also have a
pairwise form. For the antisymmetric color structure, the result
is~\cite{Ferroglia:2009ii}
\begin{align}
\label{eq:planarCTfeyngauge}
F^{(b)}_\mathrm{Feyn.}+F^{(c)}_\mathrm{Feyn.} &= \frac{1}{2} (\g_{ij}\coth \g_{ij})\x\coth\g_{jk}\p{\g_{jk}^2+2\g_{jk}\log(1-e^{-2\g_{jk}})-\Li_2(e^{-2\g_{jk}})+\frac{\pi^2}{6}}\nn\\
&\qquad+\,\textrm{antisym.},
\end{align}
When all the Wilson lines are lightlike, the sum of graphs actually vanishes in
Feynman gauge, a result which is not immediately obvious.  At large $\gamma$,
the $-\frac{1}{2}\gamma_{ij}\g_{jk}^2$ asymptotic behavior of the non-planar
amplitude in \eq{nonplanarfeyngauge} is exactly canceled by contributions from
\eq{planarCTfeyngauge}.

Returning to the difficult non-planar graph, the simplicity of the final
result $F_{\mathrm{Feyn}}^{(a)}$ is easily understood from the AdS picture.
In $\R\x\AdS$, each Wilson line points in the time direction, and thus sources only the $\tau$ component of the gauge field.
In conformal gauge $A_\tau$ does not mix with $A_i$, then the interaction vertex in \ref{fig:2loop-nonplanar}
 involves three $\tau$-polarized gauge fields, and thus vanishes identically.
In this gauge, only the planar and counterterm diagrams contribute, and $F$ naturally has a pairwise-factorized form.
 Note that one of the factors $(\g_{ji}\coth\g_{ij})$ looks just like $\G_\mathrm{cusp}^\mathrm{1-loop}$.  This is suggestive that the contribution of $F_{\mathrm{Feyn}}^{(a)}$ should come from the counterterm graph in conformal gauge.
We will shortly show that this is indeed the case.

Let us rephrase the above argument in a way that emphasizes the analogy with our
one-loop computation in Section~\ref{sec:ClassicalAdSEnergies}.  Notice that
each line in~\ref{fig:2loop-nonplanar}, thought of as a static charge in $\AdS$,
couples only to the $\tau$-independent modes of the gauge field in
$\R\x\AdS$.\footnote{This is not the case for diagrams involving more than one
  gluon emission from a single Wilson line.}  Thus, the computation of this
graph ``dimensionally reduces" from $\R\x\AdS$ to $\AdS$, with one overall
$\tau$-integral contributing a logarithmic divergence multiplying the anomalous
dimension.  The theory on $\AdS$ contains a scalar (coming from $A_\tau$), and a
three-dimensional gauge field.  Each Wilson line sources the scalar, so
diagram~\ref{fig:2loop-nonplanar} becomes a three-point function of scalars in
$\AdS$.  Finally, since there is no three-scalar interaction in the dimensional
reduction of Yang-Mills theory, this correlator vanishes at leading order.  The
utility of conformal gauge is that it makes dimensional reduction in the
$\tau$-direction much simpler than it would be in Feynman gauge.

Having understood why $F(\g_{ij},\g_{jk},\g_{ki})$ should have a simple form,
let us proceed to compute it using conformal gauge.
Diagram~\ref{fig:2loop-nonplanar} now vanishes, and the entire contribution
comes from the planar and counterterm graphs.  All Wilson lines point in the
$\tau$-direction, so we need only the radial part of the gauge-boson propagator
which from Eq.~(\ref{eq:propangularradial}) is:
\begin{equation}
\label{eq:radialAdSprop}
D^{(\tau\tau)}_{\mu\nu}(x,y) = -\ka_d\frac{x_\mu y_\nu}{x^2y^2}\p{\frac{x\.y}{[-(x-y)^2]^{d/2-1}}-[-(x-y)^2]^{2-d/2}
  +\chi\big(|x|,|y|\big)^{4-d}}.
\end{equation}
The first term in parentheses is the same as $D^F_{\tau\tau}$ in Feynman gauge. The second and third terms are new.

Since we seek the coefficient of the antisymmetric color structure $f^{abc}T^a_i T^b_j T^c_k$, which does not arise at one-loop, the anomalous dimension is simply the coefficient of the $1/\e$ pole in the sum of diagrams~\ref{fig:2loop-planar} and \ref{fig:2loop-ct}.  To separate UV and IR divergences, we must in general regulate the IR with something other than dimensional regularization.
 However, here we can safely ignore this subtlety since the entire divergence structure comes from a single scaleless integral, and it will be simple to isolate the associated UV divergence.  We have checked that a more careful treatment of the infrared, for instance giving the Wilson lines some finite length, yields the same results.

To evaluate \ref{fig:2loop-planar}, let us first perform the integrals along the lines with only a single gluon emission.
 This gives the Coulomb potential from a
single Wilson line, which we now need up to order $\e$. We have, with the normalization $n_1^2=n_2^2=1$,
\begin{align}
\label{eq:oneintegral}
    \int_0^\infty ds\,n_i^\mu n_j^\nu D_{\mu\nu}(s n_i,t n_j)
    &=\kappa_d(-1)^\epsilon\int_0^\infty \rd s\left(\frac{-n_i\cdot n_j}{(sn_i-tn_j)^{2-2\epsilon}}
    +\frac{\chi(s,t)^{2\epsilon}-(sn_i-tn_j)^{2\epsilon}}{st}\right)\nn\\
    &=\kappa_d\frac{(-1)^\epsilon}{t^{1-2\epsilon}}\p{ E_F^{(0)}(\g_{ij}) + \e E_F^{(1)}(\g_{ij}) +\e E_C^{(1)}(\g_{ij})}
\end{align}
Here $E_F^{(0)}=A_\tau(\gamma_{ij})$ from Eq.~\eqref{eq:Atau} is just the scalar potential from a Wilson line
in 4 dimensions:
\begin{equation}
  E_F^{(0)}(\g) = \gamma \coth \gamma
\end{equation}
 $E_F^{(1)}$ is the next term in the $\e$ expansion of this potential in Feynman gauge,
also coming from the first term in the integral. It is
\begin{align}
    E_F^{(1)}(\g)=\coth\g\p{\g^2+2\g\log(1-e^{-2\g})-\Li_2(e^{-2\g})+\frac{\pi^2}{6}}
\end{align}
Finally, $E_C^{(1)}$ is the new piece present in conformal gauge and not in
Feynman gauge, from the second term in the integral in \eq{oneintegral}.  It gives
\begin{align}
    E_C^{(1)}(\gamma_{ij})
    &=\int_0^\infty\frac{\rd s}{s}\log\frac{\chi(s,1)^2}{(n_is-n_j)^2}\\
    &=\int_{-\infty}^\infty \rd\tau\log\Big(\frac{\cosh\tau}{\cosh\tau+\cosh\gamma_{ij}}\Big) +\int_0^\infty\frac{\rd s}{s}
 \log\frac{\chi(s,1)^2}{1+s^2}
 \nn \\
     &= -\gamma_{ij}^2 -\frac{\pi^2}{4} + c_\chi \,.
  \nn
\end{align}
The constant $c_\chi$ is a gauge-dependent but $\gamma$-independent number which
will cancel from the final result (and is exactly zero for
$\chi(|x|,|y|)=\sqrt{x^2+y^2}$).  Note that the asymptotic expansion of
$E_C^{(1)}$ at large $\gamma$ is $-\gamma^2$, which cancels the asymptotic
expansion of $E_F^{(1)}$, leaving zero contribution to the antisymmetric color
structure in the anomalous dimension for the light-like limit. Two loop graphs
involving only two lines do contribute in the light-like limit, and give an
energy which grows linearly with the cusp angle.

With $\mathcal{O}(\e)$ parts of the scalar potential calculated, it is now easy to extract the antisymmetric
part of the two-loop anomalous dimension. The counterterm and planar  graphs can be combined into
\begin{align}
I^{(b)}+I^{(c)}=&\int_0^\oo \frac{\rd t_1}{t_1^{1-2\e}}\Big[
E_F^{(0)}(\gamma_{ij}) + \e E_F^{(1)}(\gamma_{ij})+\e
E_C^{(1)}(\gamma_{ij})\Big]
 \\
&\times\left\{-\frac{1}{\e} E_F^{(0)}(\gamma_{jk}) +
\int_0^{t_1} \frac{\rd t_2}{t_2^{1-2\e}}\Big[ E_F^{(0)}(\gamma_{jk}) + \e E_F^{(1)}(\gamma_{jk})+\e E_C^{(1)}(\gamma_{jk})\Big]
\right\}+ \textrm{antisym.} \nn
\end{align}
After antisymmetrizing, everything vanishes except for the cross term between
the counterterm and the $\e$ terms on the first line.  Replacing the scaleless
$t_1$ integral on the first line with $\frac{1}{2\e}$ as before, we see that
these graphs sum to produce a contribution to the anomalous dimension of the
form
\begin{align}
F^{(b)}+F^{(c)} &= \frac{1}{2} E_F^{(0)}(\g_{ij})
\left( E_F^{(1)}(\g_{jk})+ E_C^{(1)}(\g_{jk})\right)+ \textrm{antisym.},
  \nn \\
&= \frac{1}{2}\g_{ij} \coth \g_{ij}\coth \g_{jk}  \left(\g_{jk}^2+2\g_{jk}\log(1-e^{-2\g_{jk}})-\Li_2(e^{-2\g_{jk}})+\frac{\pi^2}{6}\right) \nn\\
&\qquad \qquad -\frac{1}{2}\g_{ij} \coth \g_{ij} \g_{jk}^2 + \textrm{antisym.}
\end{align}
which precisely matches $F^{(a)}+F^{(b)}+F^{(c)}$ in Feynman gauge. The
difficult non-planar graph was reproduced with a far simpler calculation
involving the $\mathcal{O}(\e)$ parts of the conformal gauge propagator.  Note
that the gauge-dependent constant $c_\chi$ drops out due to the
antisymmetrization.

For three light-like Wilson lines the calculation of the diagrams in
\fig{2loopgraphs} can also be considered directly using a conformal gauge. To do
this we should use the conformal gauge from \eq{lightlikeconformal}, rather than
the one in \eq{propangularradial}. This conformal gauge has no $D_{\tau i}$ or
$D_{i\tau}$ mixing terms and is identical to Feynman gauge for $D_{\tau\tau}$ in
$d$-dimensions.  The lack of mixing terms immediately implies that
\fig{2loopgraphs}a is zero, a result that is only seen in Feynman gauge by
direct computation~\cite{Aybat:2006wq}. For the remaining diagrams,
\fig{2loopgraphs}b and \fig{2loopgraphs}c, the calculation is identical to the one
in Feynman gauge, so the sum of these diagrams is zero just as it is
there~\cite{Aybat:2006wq,Aybat:2006mz}.

\section{Relation to Witten Diagrams in the Lightlike Limit}
\label{sec:wittendiagrams}

\begin{figure}
\begin{center}
\begin{psfrags}
\psfrag{r}[B][B][1][0]{$\R\x\AdS$}
\psfrag{a}[B][B][1][0]{$\AdS$}
\psfrag{L}[B][B][1][0]{$n_i^2\to 0$}
\includegraphics[width=\textwidth]{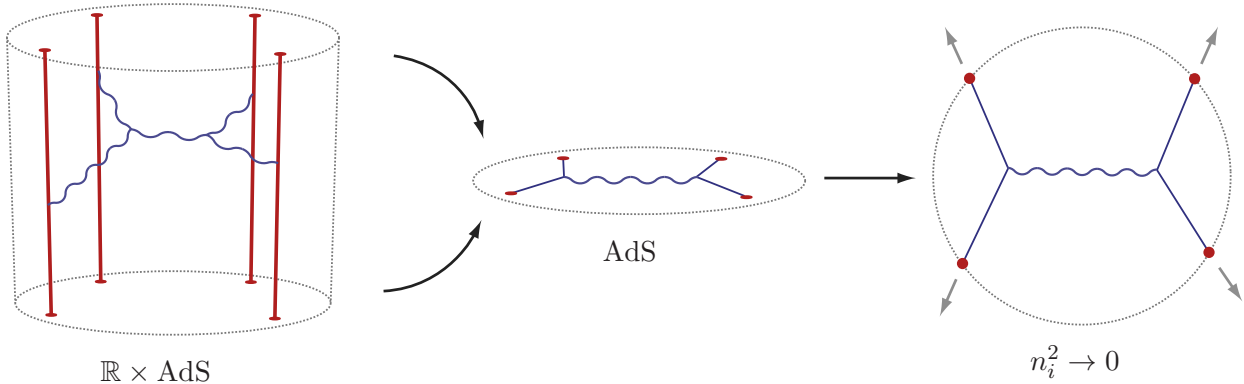}
\end{psfrags}
\end{center}
\caption{After doing all the $\tau$-integrals, the computation of certain diagrams dimensionally reduces from $\R\x\AdS$ to $\AdS$.  In the limit where the charges approach the boundary $n_i^2\to 0$, we obtain a Witten diagram.}
\label{fig:witten}
\end{figure}

Finally, let us comment on an interesting formal similarity between the
perturbation expansion for $\G_\mathrm{cusp}(n_i)$ in the lightlike limit and
the Witten diagram expansion for AdS scattering amplitudes, which has been well
studied in the AdS/CFT literature
\cite{Witten:1998qj,Freedman:1998tz,Freedman:1998bj,DHoker:1998gd,DHoker:1999jc,DHoker:1999ni}.
Recall from the previous section that diagrams with at most one gluon attached
to each line involve only $\tau$-independent modes of the gauge field.  After
performing integrals in the $\tau$-direction, they become AdS scattering
amplitudes in a gauge theory containing an adjoint scalar, which is sourced by
each charge.  As the parton directions become lightlike $n_i^2\to 0$, the
corresponding charges move off to the boundary of $\AdS$.  We are left, at least
formally, with a boundary-to-boundary scattering amplitude --- a Witten diagram
(Figure \ref{fig:witten}).

We could have anticipated some relation to Witten diagrams simply from the geometry.  Our realization of $\AdS_3$ as a hyperboloid inside $\R^{1,3}$ is known in the AdS/CFT literature as the {\it embedding space} (or sometimes {\it covering space}) formalism \cite{Dirac:1936fq,Mack1969174,Boulware:1970ty,Ferrara:1973yt,Weinberg:2010fx,Costa:2011mg}.  Its utility is that the isometries of AdS (and conformal transformations of its boundary) become linearly-realized Lorentz transformations of the embedding space, an observation that dates back to Dirac \cite{Dirac:1936fq}.  Here, we arrived at this formalism from the other direction, beginning in Minkowski space, and finding that $\AdS$ geometry appears naturally.

A possible correspondence between Feynman diagrams for $\G_\mathrm{cusp}(n_i)$
and Witten diagrams is complicated by several issues.  Firstly, as we argued
extensively in Section~\ref{sec:ClassicalAdSEnergies}, choosing proper boundary
conditions in $\AdS$ is crucial for capturing the correct physics.  For example, our
scalar bulk-to-bulk propagator is the potential we computed in
Eq.~(\ref{eq:2Jenergy}),
\begin{equation}
D_{BB}(n_1,n_2) = \frac 1 {4\pi^2}\p{\pi+i\b_{12}} \coth \b_{12}.
\end{equation}
Recall that this solves Laplace's equation in the presence of a constant charge
density, so does not correspond to the usual bulk-to-bulk scalar propagator,
which solves the homogeneous Laplace's equation in AdS.  In the limit that one
of the directions $n_1$ becomes lightlike, we also obtain an unorthodox
boundary-to-bulk propagator,
\begin{equation}
\label{eq:boundarytobulk}
D_{\ptl B}(n_1,n_2) = \lim_{|n_1|\to 0}D_{BB}(n_1,n_2) = \lim_{|n_1|\to 0}\frac{i}{4\pi^2} \p{-\log |n_1|+\log\frac{n_1\.n_2}{|n_2|}}
\end{equation}
This is divergent as $|n_1|\to 0$, a reflection of the collinear singularities that arise in this limit.\footnote{Its form is perhaps reminiscent of the traditional boundary-to-bulk propagator for a scalar with an infinitesimal mass
\begin{equation}
\label{eq:boundarytobulkspeculative}
D_{\ptl B}(n_1,n_2)\stackrel{?}{\sim} \lim_{\De\to 0}\frac{i}{4\pi^2}\frac 1 {\De}\p{\frac{n_1\.n_2}{|n_2|}}^{\De}
\end{equation}
where the mass is related to $\De$ via the usual AdS/CFT dictionary, $m^2R_\AdS^2 = \De(\De-2)$.  One might speculate that in a calculation where the $-\log|n_i|$ singularities are regulated and cancel appropriately, Eq.~(\ref{eq:boundarytobulkspeculative}) might be an acceptable form for the boundary-to-bulk propagator.}

Contributions to $\G$ also differ from traditional Witten diagrams in their
contours of integration.  For Witten diagrams, one performs bulk integrals over
$\AdS$, whereas Feynman integrals involve all of Minkowski space, so should
incorporate both copies of $\AdS$ and dS as well (Figure~\ref{fig:copiesofAdS}).

To use Witten diagrams for computing $\G$, one must account for all of the above
issues.  Nevertheless, the possible applicability of AdS amplitude technology is
encouraging.  Witten diagrams have been well-studied, and recently new
techniques involving Mellin representations have substantially improved
efficiency of computation
\cite{Mack:2009mi,Mack:2009gy,Penedones:2010ue,Paulos:2011ie,Fitzpatrick:2011ia}.

As an example, the gauge-boson exchange diagram depicted in
Figure~\ref{fig:witten} was first computed with traditional boundary conditions
over a decade ago \cite{DHoker:1998gd,DHoker:1999jc,DHoker:1999ni}.  The answer
is a nontrivial sum of so-called $D$-functions
$D_{\De_1,\De_2,\De_3,\De_4}(u,v)$, where $u$ and $v$ are conformal
cross-ratios.  Formally, this diagram contributes to the regular anomalous
dimension $\g(\a_s,\{n_i\.n_j\})$ in Eq.~(\ref{eq:neubertgardiconjecture}) at
three-loops, with a color structure $f^{abe}f^{cde}\bm T_i^a\bm T_j^b \bm T_k^c
\bm T_l^d$.  While we have understood using conformal gauge why
$\G_\mathrm{cusp}(n_i)$ should have a pairwise structure up to two-loops, we see
no {\it a priori} reason that this structure should persist to higher orders.  We
interpret the fact that the diagram in Figure~\ref{fig:witten} is nonzero as an
indication that $\g(\a_s,\{n_i\.n_j\})$ might very well have nontrivial
conformal cross-ratio dependence. Very recently, two papers have
appeared~\cite{DelDuca:2011ae,DelDuca:2011xm} which propose that additional
constraints on the conformal cross ratio dependence of the soft anomalous
dimension follow from consideration of the Regge limit. Perhaps by using
tools developed for Witten diagrams, the 3-loop anomalous dimension can be
calculated exactly, hopefully resolving the controversy in \fact 6.

\section{Conclusions} \label{sec:conclusion}

In this paper we have discussed how properties of operators $\cW$ built from $N$ Wilson lines
can be understood in radial coordinates. These operators
appear in high energy collisions that produce jets, where the lines extend out
from the location of the hard interaction, taken to be the origin. In
radial coordinates, $\R\x\AdS$, the direction of the Wilson lines are specified by points
in Euclidean $\AdS_3$ and motion along any of the Wilson lines corresponds to
time-translations of $\tau\in\R$.

We have demonstrated that many of the key properties of anomalous dimensions of
these operators have an intuitive and simple description in terms of these
static charges in $\AdS$. In particular: the dependence on cusp angles
$\beta_{ij}$ just corresponds to the geometric distance between the lines in
$\AdS$; the one-loop anomalous dimension of $\cW$ is given by a classical energy
computation on $\AdS$ (with special care given to boundary conditions). 

There is
an intuitive physical picture associated to the real and imaginary parts of the
anomalous dimension.  This picture leads to an intriguing analogy between the at
most linear growth of imaginary energy with separation in AdS, guaranteed by the
at most linear growth of the anomalous dimensions with cusp angles, and the linear
dependence of energy on separation for charges in gauge theories, which is
associated with the flux-tube picture of confinement.

To fully exploit the physical picture arising in $\R\x\AdS$ we introduced a
class of gauges, referred to as conformal gauges. In these gauges, there is no
kinetic mixing between temporal components of the gauge field, $A_\tau$, and
spatial components, $A_i$. Conformal gauges are formulated in position space in
$d$-dimensions, so that they are suitable for calculations using dimensional
regularization. Conformal gauges in $\R\x\AdS$ are effectively the analog of Feynman
gauge in flat space, and simplify some perturbative computations involving
Wilson lines. Since all Wilson lines are only sources for $A_\tau$, the absence
of mixing directly implies that one only has to consider scalar exchange at
leading orders in perturbation theory.  In particular graphs involving three
gluon or four gluon vertices may vanish simply from the absence of $(A_\tau)^3$
and $(A_\tau)^4$ interactions in QCD. We have demonstrated this explicitly by
considering a two-loop computation involving three time-like Wilson lines, and
showing that it reduces to a one-loop computation with counterterm insertions.
For three light-like Wilson lines we have also shown that a suitable conformal
gauge simplifies this calculation by making it explicit that the most
complicated diagram involving the three-gluon vertex vanishes.

Many avenues remain open to future exploration, and we have only briefly touched
on a few of them. In the limit where one or more Wilson lines become light-like,
extra ultraviolet and infrared divergences appear, and new features emerge in
the anomalous dimension of $\cW$, such as dependence on the renormalization
group scale $\mu$. While we have formulated a suitable conformal gauge for use
with light-like lines, we have not explored in detail many interesting
computations, such as the two-loop anomalous dimension from two light-like
lines, or graphs occurring in soft functions that have real radiation.  Many
interesting questions only appear for $\cW$ with four lines taken at three loops
and beyond, such as possible dependence of the anomalous dimension on conformal
cross-ratios.  We anticipate that the use of conformal gauges will be a powerful
technique for analyses which seek to definitively answer questions which appear
at this order.

We have also observed a relationship between diagrams with multiple Wilson lines and Witten diagrams, which have
been studied extensively in the context of the AdS/CFT correspondence. There is hope that technology developed for
computing these Witten diagrams can be used directly for calculations about Wilson lines, with direct application
to jet physics, and possibly also to improved understanding of the structure of amplitudes in gauge theories.

\section*{ACKNOWLEDGMENTS}

This work was supported in part by the Offices of High Energy and Nuclear
Physics of the U.S.\ Department of Energy under the Contracts DE-SC003916 and
DE-FG02-94ER40818, and by the Alexander von Humboldt foundation.  The authors
thank J.~Maldacena and D.~Hofman for many helpful discussions about the AdS
picture, and T.~Becher and A.~Manohar for helpful comments.  MDS would like to
thank the KITP for hospitality, and its participants including E.~Gardi and
M.~Neubert for helpful discussions. DSD would also like to thank C.~C\'ordova,
F.~Denef, R.~Loganayagam, and D.~Poland for discussions.

\appendix

\section{General Class of Conformal Gauges}
\label{app:genconformal}

The most general possible form of the gauge transformation one-form
is\footnote{More general forms are possible if we introduce one or more
  additional fixed vectors in the gauge transformation, such as a $v^\mu$ where
  $v^2=1$.}
\begin{equation}
\L_\mu(y,x) = \frac{\ka_d}{(|x||y|)^{d/2-1}} \big[ x_\mu g_1(\a,\b) + y_\mu
g_2(\a,\b) \big]
,\qquad\textrm{where}\quad \a\equiv\frac{x\.y}{|x||y|},\ \b\equiv\frac{|y|}{|x|},
\end{equation}
and $g_{1,2}(\a,\b)$ are functions to be specified. Eq.~(\ref{eq:Dmunu}) yields
the propagator
\begin{align} \label{eq:generalD}
  D_{\mu\nu}(x,y)
  =  \frac{\ka_d}{(|x||y|)^{d/2-1}} \Big[
  g_{\mu\nu} A  + \frac{x_\mu x_\nu}{x^2} B + \frac{y_\mu y_\nu}{y^2} C
   + \frac{x_\mu y_\nu}{|x||y|} E + \frac{y_\mu x_\nu}{x\cdot y} Z  \Big] ,
\end{align}
where   we have
\begin{align} \label{eq:ABCEZ}
  A &= -[2\a -\b -\b^{-1}]^{1-d/2} + g_2 + \bar g_2
   \,, \nn\\
  B & = \b^{-1} g_1^{(1,0)}+ (1-d/2)\, \bar g_2 -\a\, \bar
  g_2^{(1,0)} + \b^{-1}\, \bar g_2^{(0,1)}
   \,, \nn\\
  C & = \b\, \bar g_1^{(1,0)}+ (1-d/2)\,  g_2 -\a\,
  g_2^{(1,0)} + \b\,  g_2^{(0,1)}
   \,, \nn\\
  E & = g_1^{(0,1)} + \bar g_1^{(0,1)} + (1-d/2)(\b^{-1} g_1 + \b\, \bar g_1)
       - \a \big( \b^{-1} g_1^{(1,0)} + \b\, \bar g_1^{(1,0)} \big)
   \,, \nn\\
  Z & = \a g_2^{(1,0)} + \a \bar g_2^{(1,0)}
  \,,
\end{align}
with the definitions $\bar g_i(\a,\b)=g_i(\a,\b^{-1})$, $g_i^{(1,0)} = \partial
g_i(\a,\b)/\partial\a$, $g_i^{(0,1)}=\partial g_i(\a,\b)/\partial\b$, $\bar
g_i^{(0,1)}=\partial g_i(\a,\b^{-1})/\partial\b^{-1}$, etc.  The conformal gauge
conditions, which ensure there is no mixing between time and spatial directions
in $\R\x\AdS$, are $x^\mu D_{\mu\nu}(x,y) A^\nu(y)=0$ when $y_\nu A^\nu(y)=0$,
and $A^\mu(x) D_{\mu\nu}(x,y) y^\nu=0$ when $x_\mu A^\mu(x)=0$. These require
\begin{align} \label{eq:generalnomix}
  A + B + Z = 0 \,,\qquad  B=C \,,
\end{align}
which are two differential equations for the functions $g_1$ and $g_2$.
Substituting \eq{generalnomix} into \eq{generalD} yields the general result for
the conformal gauge propagator
\begin{align} \label{eq:cgaugeD}
  \tilde D_{\mu\nu}(x,y) &= \frac{\ka_d}{(|x||y|)^{d/2-1}} \bigg[
   \Big( g_{\mu\nu} - \frac{x_\mu x_\nu}{x^2} - \frac{y_\mu
    y_\nu}{y^2} + \frac{x\cdot y\, x_\mu y_\nu}{x^2 y^2}\Big) A
 \\
   & \qquad
  + \Big(\frac{y_\mu x_\nu}{x\cdot y}  - \frac{x_\mu x_\nu}{x^2} - \frac{y_\mu
    y_\nu}{y^2} + \frac{x\cdot y\, x_\mu y_\nu}{x^2 y^2}\Big) Z
   + \frac{x\cdot y\, x_\mu y_\nu}{x^2y^2} \Big( \frac{E}{\a} - A - Z\Big)
  \bigg]
\,. \nn
\end{align}
The first two tensor structures are spatial (angular), while the latter is
temporal (radial).  Using \eq{generalnomix} and \eq{ABCEZ} we can write
\begin{align}
  \frac{E}{\alpha} - A - Z &= - (2\a -\b-\b^{-1})^{1-d/2} + \Big[ (2-d/2)(g_2+\bar
  g_2) + \b\, g_2^{(0,1)}+\b^{-1} \bar g_2^{(0,1)} \Big]
 \nn\\
 & + \a^{-1} \Big[  (1-d/2)(\b^{-1} g_1 +\b\, \bar g_1) +  g_1^{(0,1)}
   +  \bar g_1^{(0,1)} \Big] \,,
\end{align}
where the first term is the result from Feynman gauge and the last two terms are
induced by the gauge transformation.

To consider the light-like limit for $x$ and $y$ we take $x^\mu = \lambda_1
n_1^\mu$ and $y^\mu = \lambda_2 n_2^\mu$, where without loss of generality we
take $\lambda_i>0$ and send $n_1^2=n_2^2\to 0$.  This leaves
$\beta=\lambda_2/\lambda_1$ fixed and sends $\alpha\to \infty$. For the propagator
between points on two light-like Wilson lines only the last term in \eq{cgaugeD}
contributes,
\begin{align} \label{eq:lightlikeD}
  n_1^\mu n_2^\nu \tilde D_{\mu\nu}(x,y) &=
  \kappa_d\,n_1\cdot n_2
  \lim_{n_i^2\to \infty}
   (\lambda_1^2 \lambda_2^2 n_1^2n_2^2)^{1/2-d/4}
    \bigg[ \frac{E}{\alpha} - A - Z \bigg] \,.
\end{align}
To ensure this gives the same result as Feynman gauge we can choose a conformal
gauge where
\begin{align} \label{eq:lightlikeconformal}
g_1(\a,\b) = -\a\b\, g_2(\a,\b) \,,
\end{align}
which makes $E/\a-A-Z = - (2\a -\b-\b^{-1})^{1-d/2}$.
Equation~(\ref{eq:lightlikeD}) then becomes
\begin{align}
  n_1^\mu n_2^\nu \tilde D_{\mu\nu}(x,y) &= -\kappa_d\, n_1\cdot n_2 (2 n_1\cdot
  n_2 \lambda_1 \lambda_2)^{1-d/2} \,,
\end{align}
which is the same as the Feynman gauge result. The $A+B+Z=0$ and $B=C$ no mixing
conditions for this case becomes
\begin{align}
  \b\, g_2^{(0,1)} + (2-d/2) g_2 = (2\a -\b -\b^{-1})^{1-d/2}\,,
\end{align}
which implies
\begin{align} \label{eq:g2new}
  g_2(\a,\b) = \b^{d/2-2} (\a^2-1)^{1-d/2}\, (\b-\a)\:
  {}_2F_1\Big(\frac12, \frac{d}{2}-1,\frac32,\frac{(\a-\b)^2}{\a^2-1}\Big) +
   \b^{d/2-2} h(\alpha)
  \,,
\end{align}
with an arbitrary function $h(\alpha)$ that still must be fixed to fully specify
the gauge. Using Eq.~(\ref{eq:ABCEZ}),
Eqs.~(\ref{eq:lightlikeconformal}) and (\ref{eq:g2new}) determine the spatial
terms in the conformal gauge boson propagator in \eq{cgaugeD}. It is
straightforward to verify that the propagator is non-singular in the limit
$d\to 4$.

\section{Ghosts in Conformal Gauge} \label{app:ghost}

In position space the Feynman rules for ghosts are more easily represented with
a ``ghost field'' $G_\mu^{abc}$ which is the product of a ghost propagator and
ghost-gluon vertex. For the gauge transformation in \eq{Dmunu} the appropriate
ghost field is~\cite{Cheng:1986hu}
\begin{align}
 G_\mu^{abc}(y,x) = -i g f^{abc} \big[ \partial_y^\nu D_{\mu\nu}^F(x,y) +
 (\partial_\mu^x \partial_\nu^x - g_{\mu\nu} \Box_x) \Lambda^\nu(y,x) \big] \,,
\end{align}
where $D_{\mu\nu}^F$ is the Feynman gauge gluon propagator from Eq.~(\ref{DFd}) and
$\Lambda^\nu(y,x)$ is the one-form appearing in the gauge transformed gluon
propagator $D_{\mu\nu}$. This result can be used for any member of the general
class of conformal gauges discussed in \app{genconformal}.

\bibliography{conformal}

\end{document}